\documentclass[reprint,superscriptaddress,preprintnumbers,longbibliography,
amsmath,amssymb,aps,superscriptaddress,floatfix,tightenlines,prl
]{revtex4-2}
\usepackage{subfigure}
\usepackage{mathrsfs}
\usepackage[T1]{fontenc}
\usepackage{graphicx}
\usepackage{dcolumn}
\usepackage{bm}
\usepackage{color}
\usepackage{times}
\usepackage[colorlinks=true,
citecolor=magenta,
linkcolor=red,
anchorcolor=black,
urlcolor=purple]{hyperref}
\usepackage{overpic}
\usepackage{braket}
\usepackage{changes}
\definechangesauthor[name={syh},color=red]{syh}

\begin{document}
\title{Direct Implementation of High-Fidelity Three-Qubit Gates for Superconducting Processor with Tunable Couplers}

\author{Hao-Tian Liu}
	\thanks{These authors contributed equally to this work.}
    \affiliation{Beijing National Laboratory for Condensed Matter Physics, Institute of Physics, Chinese Academy of Sciences, Beijing 100190, China}
	\affiliation{School of Physical Sciences, University of Chinese Academy of Sciences, Beijing 100049, China}
	
\author{Bing-Jie Chen}
	\thanks{These authors contributed equally to this work.}
    \affiliation{Beijing National Laboratory for Condensed Matter Physics, Institute of Physics, Chinese Academy of Sciences, Beijing 100190, China}
	\affiliation{School of Physical Sciences, University of Chinese Academy of Sciences, Beijing 100049, China}
	
\author{Jia-Chi Zhang}
    \affiliation{Beijing National Laboratory for Condensed Matter Physics, Institute of Physics, Chinese Academy of Sciences, Beijing 100190, China}
	\affiliation{School of Physical Sciences, University of Chinese Academy of Sciences, Beijing 100049, China}
	
\author{Yong-Xi Xiao}
    \affiliation{Beijing National Laboratory for Condensed Matter Physics, Institute of Physics, Chinese Academy of Sciences, Beijing 100190, China}
	\affiliation{School of Physical Sciences, University of Chinese Academy of Sciences, Beijing 100049, China}	
	
\author{Tian-Ming Li}
    \affiliation{Beijing National Laboratory for Condensed Matter Physics, Institute of Physics, Chinese Academy of Sciences, Beijing 100190, China}
	\affiliation{School of Physical Sciences, University of Chinese Academy of Sciences, Beijing 100049, China}		
	
\author{Kaixuan Huang}
    \affiliation{Beijing Key Laboratory of Fault-Tolerant Quantum Computing, Beijing Academy of Quantum Information Sciences, Beijing 100193, China}

\author{Ziting Wang}
    \affiliation{Beijing Key Laboratory of Fault-Tolerant Quantum Computing, Beijing Academy of Quantum Information Sciences, Beijing 100193, China}

\author{Hao Li}
    \affiliation{Beijing Key Laboratory of Fault-Tolerant Quantum Computing, Beijing Academy of Quantum Information Sciences, Beijing 100193, China}
    
\author{Kui Zhao}
    \affiliation{Beijing Key Laboratory of Fault-Tolerant Quantum Computing, Beijing Academy of Quantum Information Sciences, Beijing 100193, China}
    
\author{Yueshan Xu}
    \affiliation{Beijing Key Laboratory of Fault-Tolerant Quantum Computing, Beijing Academy of Quantum Information Sciences, Beijing 100193, China}
    
\author{Cheng-Lin Deng}
    \affiliation{Beijing National Laboratory for Condensed Matter Physics, Institute of Physics, Chinese Academy of Sciences, Beijing 100190, China}
	\affiliation{School of Physical Sciences, University of Chinese Academy of Sciences, Beijing 100049, China}	
	
\author{Gui-Han Liang}
    \affiliation{Beijing National Laboratory for Condensed Matter Physics, Institute of Physics, Chinese Academy of Sciences, Beijing 100190, China}
	\affiliation{School of Physical Sciences, University of Chinese Academy of Sciences, Beijing 100049, China}		
	
\author{Zheng-He Liu}
    \affiliation{Beijing National Laboratory for Condensed Matter Physics, Institute of Physics, Chinese Academy of Sciences, Beijing 100190, China}
	\affiliation{School of Physical Sciences, University of Chinese Academy of Sciences, Beijing 100049, China}		
	
\author{Si-Yun Zhou}
    \affiliation{Beijing National Laboratory for Condensed Matter Physics, Institute of Physics, Chinese Academy of Sciences, Beijing 100190, China}
	\affiliation{School of Physical Sciences, University of Chinese Academy of Sciences, Beijing 100049, China}		
	
\author{Cai-Ping Fang}
    \affiliation{Beijing National Laboratory for Condensed Matter Physics, Institute of Physics, Chinese Academy of Sciences, Beijing 100190, China}
	\affiliation{School of Physical Sciences, University of Chinese Academy of Sciences, Beijing 100049, China}		
	
\author{Xiaohui Song}
    \affiliation{Beijing National Laboratory for Condensed Matter Physics, Institute of Physics, Chinese Academy of Sciences, Beijing 100190, China}
    \affiliation{School of Physical Sciences, University of Chinese Academy of Sciences, Beijing 100049, China}
    \affiliation{Hefei National Laboratory, Hefei 230088, China}
    
\author{Zhongcheng Xiang}
    \affiliation{Beijing National Laboratory for Condensed Matter Physics, Institute of Physics, Chinese Academy of Sciences, Beijing 100190, China}
    \affiliation{School of Physical Sciences, University of Chinese Academy of Sciences, Beijing 100049, China}	
    \affiliation{Hefei National Laboratory, Hefei 230088, China}
    
\author{Dongning Zheng}
    \affiliation{Beijing National Laboratory for Condensed Matter Physics, Institute of Physics, Chinese Academy of Sciences, Beijing 100190, China}
    \affiliation{School of Physical Sciences, University of Chinese Academy of Sciences, Beijing 100049, China}
    \affiliation{Hefei National Laboratory, Hefei 230088, China}

\author{Yun-Hao Shi}
    \thanks{yhshi@iphy.ac.cn}
    \affiliation{Beijing National Laboratory for Condensed Matter Physics, Institute of Physics, Chinese Academy of Sciences, Beijing 100190, China}
    \affiliation{Hefei National Laboratory, Hefei 230088, China}

\author{Kai Xu}
    \thanks{kaixu@iphy.ac.cn}
    \affiliation{Beijing National Laboratory for Condensed Matter Physics, Institute of Physics, Chinese Academy of Sciences, Beijing 100190, China}
    \affiliation{School of Physical Sciences, University of Chinese Academy of Sciences, Beijing 100049, China}
    \affiliation{Beijing Key Laboratory of Fault-Tolerant Quantum Computing, Beijing Academy of Quantum Information Sciences, Beijing 100193, China}
    \affiliation{Hefei National Laboratory, Hefei 230088, China}
    \affiliation{Songshan Lake Materials Laboratory, Dongguan 523808, Guangdong, China}

\author{Heng Fan}
    \thanks{hfan@iphy.ac.cn}
    \affiliation{Beijing National Laboratory for Condensed Matter Physics, Institute of Physics, Chinese Academy of Sciences, Beijing 100190, China}
    \affiliation{School of Physical Sciences, University of Chinese Academy of Sciences, Beijing 100049, China}
    \affiliation{Beijing Key Laboratory of Fault-Tolerant Quantum Computing, Beijing Academy of Quantum Information Sciences, Beijing 100193, China}
    \affiliation{Hefei National Laboratory, Hefei 230088, China}
    \affiliation{Songshan Lake Materials Laboratory, Dongguan 523808, Guangdong, China}

\begin{abstract}
Three-qubit gates can be constructed using combinations of single-qubit and two-qubit gates, making their independent realization unnecessary. However, direct implementation of three-qubit gates reduces the depth of quantum circuits, streamlines quantum programming, and facilitates efficient circuit optimization, thereby enhancing overall performance in quantum computation.
In this work, we propose and experimentally demonstrate a high-fidelity scheme for implementing a three-qubit controlled-controlled-Z (CCZ) gate in a flip-chip superconducting quantum processor with tunable couplers.
This direct CCZ gate is implemented by simultaneously leveraging two tunable couplers interspersed between three qubits to enable three-qubit interactions, achieving an average final state fidelity of $97.94\%$ and a process fidelity of $93.54\%$. This high fidelity cannot be achieved through a simple combination of single- and two-qubit gate sequences from processors with similar performance levels. Our experiments also verify that multilayer direct implementation of the CCZ gate exhibits lower leakage compared to decomposed gate approaches. 
As a showcase, we utilize the CCZ gate as an oracle to implement the Grover search algorithm on three qubits, demonstrating high performance with the target probability amplitude significantly enhanced after two iterations. These results highlight 
the advantage of our approach, and facilitate the implementation of complex quantum circuits. 

\end{abstract}
\maketitle

\maketitle

Typically, a universal set of quantum gates for computation requires only single-qubit and two-qubit gates, from which any multiqubit gate, such as three-qubit gates, can be composed.~\cite{Barenco1995}. However, the direct construction and control of high-fidelity multiqubit gates remain crucial for advancing quantum computation~\cite{nielsen_chuang_2000}, particularly in achieving quantum error correction~\cite{Kitaev1997,toshinari_three-qubit_2024,Bluvstein2024}, quantum simulation~\cite{nguyen_programmable_2024}, and scalable quantum algorithms in the noisy intermediate-scale quantum (NISQ) era~\cite{Preskill2018}. The CCZ gate stands out as a pivotal three-qubit gate, enabling specific operations that are challenging to replicate using single- or two-qubit gates alone. This gate applies a phase shift only when all three qubits are in the target state, making it indispensable in applications like Grover's search algorithm~\cite{Grover1997,roy_programmable_2020} and quantum error correction codes~\cite{reed_realization_2012,Paetznick2013,Yoder2016,rasmussen_single-step_2020}. Direct implementation of the CCZ gate can significantly reduce circuit complexity and depth, addressing the limitations associated with decomposing complex operations into sequences of simpler gates, which introduce cumulative errors and increase operational overhead.

Despite its significance, achieving a high-fidelity CCZ gate has proven challenging across various quantum platforms, including superconducting qubits~\cite{baker_single_2022,baekkegaard_realization_2019,kim_high-fidelity_2022,warren2023extensive,nguyen_empowering_2024,glaser_controlled-controlled-phase_2023,song_continuous-variable_nodate}, trapped ions~\cite{fang_scheme_2010,monz_realization_2009}, photonic systems~\cite{wang_polarization_2024}, and cavity QED systems~\cite{chen_toffoli_2006,shi_three-qubit_2012,joshi_three-qubit_2006,moqadam_analyzing_2013}. Previous attempts have often relied on synthesizing the three-qubit gate from sequences of controlled-NOT (CNOT) gates and single-qubit rotations~\cite{wei_synthesis_2010,banchi_quantum_2016,gullans_protocol_2019,pedersen_native_2019,daraeizadeh_machine-learning_nodate,sun_quantum_2024,yu_five_2013,maslov_advantages_2016,Gu2021}, leading to complex gate sequences that extend operation times and introduce additional sources of error. Each additional gate in such sequences increases the likelihood of decoherence and operational errors, making direct implementation of the CCZ gate highly desirable for practical quantum computation. While previous works have demonstrated engineered three-body interactions through purpose-designed superconducting circuits~\cite{PhysRevLett.129.220501,nguyen2025singlestephighfidelitythreequbitgates}, these approaches often encounter inherent complexity and scalability challenges.

Experimental demonstrations of three-qubit gates in superconducting qubit systems have achieved a peak process fidelity of $98.26\%$~\cite{kim_high-fidelity_2022}. However, to accommodate the continuously expanding scale, the state-of-the-art superconducting quantum chips employ flip-chip technology and tunable coupling architectures~\cite{supremacy_Arute2019,QEC_Acharya2023,QEC_Acharya2024}. To realize high-fidelity and high-scalablity three-qubit gate operations on such multiqubit chips, we propose and experimentally demonstrate an optimized CCZ gate scheme using tunable couplers. Our approach leverages advanced fabrication techniques to directly implement high-fidelity three-qubit interactions while addressing common sources of error, such as residual two-qubit interactions and leakage to higher energy levels and couplers, through a targeted control sequence that minimizes nonadiabatic errors.

\begin{figure*}[t]
	\begin{center}
		\includegraphics[width=0.94\textwidth]{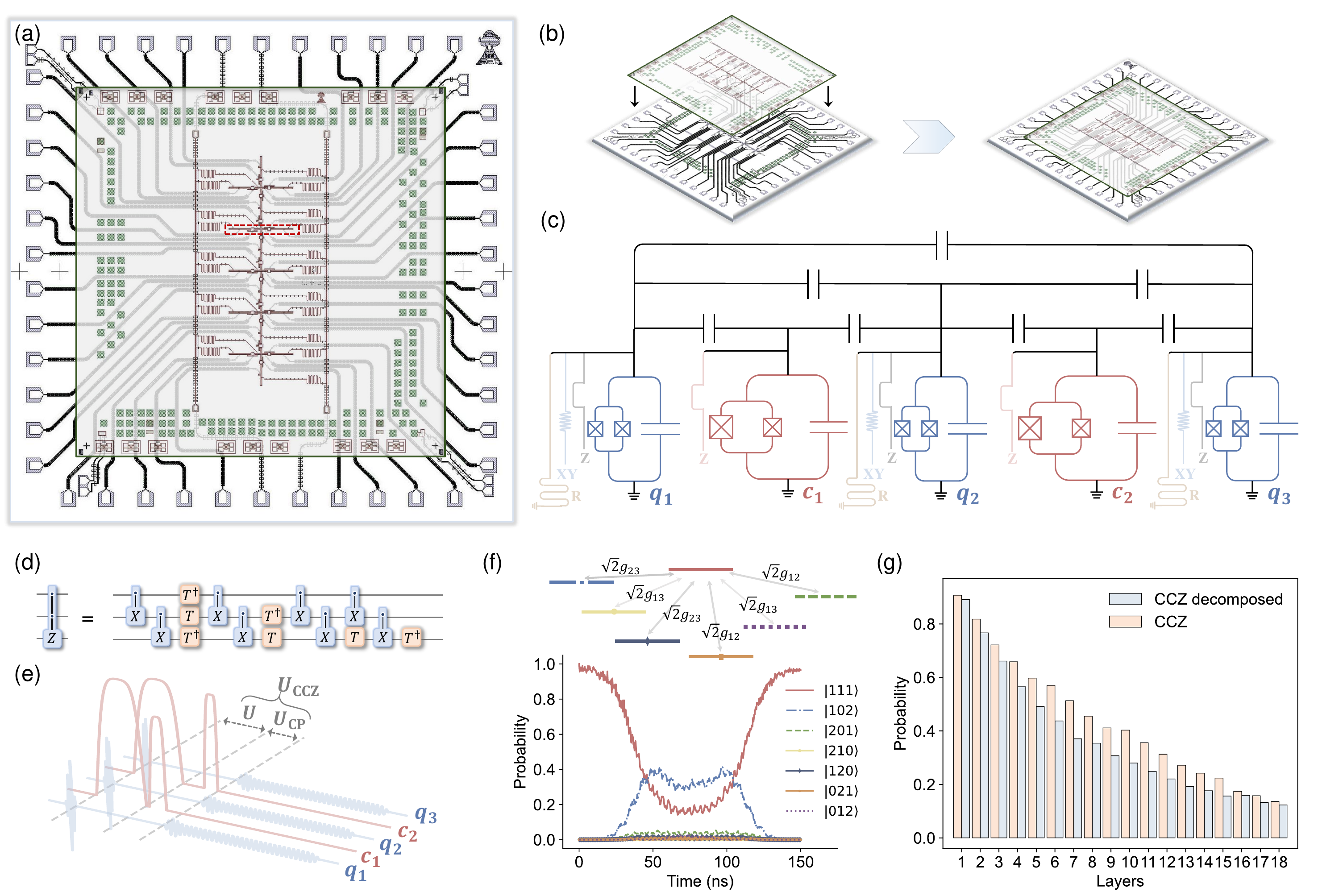}
		\caption{
		(a)~The flip-chip quantum processor with 21 superconducting qubits arranged in a 1D chain with multiple legs. Every two qubits are equipped with a coupler in between.
		(b)~Schematic of flip-chip technique. 
		(c)~Circuit diagram of the implemented superconducting circuits framed by the red dotted line in (a), consisting of three qubits ($q_1$, $q_2$, $q_3$) and two couplers ($c_1$, ${c}_2$). The qubits have the independent XY and Z controls and readout resonators, while the couplers have only Z controls.
		(d)~Quantum circuit of the CCZ gate decomposed into a series of single-qubit gates and CX (CNOT) gates.
		(e)~Pulse sequence of the direct CCZ gate for superconducting qubits with tunable couplers.
		(f)~Energy level diagram and time-dependent population transfer in the three-excitation manifold.
		(g)~Comparison of multilayer leakage between the direct CCZ gate and its decomposed implementation.
		 }\label{fig:1}
	\end{center}
\end{figure*}

\begin{figure*}[t]
	\begin{center}
		\includegraphics[width=0.91\textwidth]{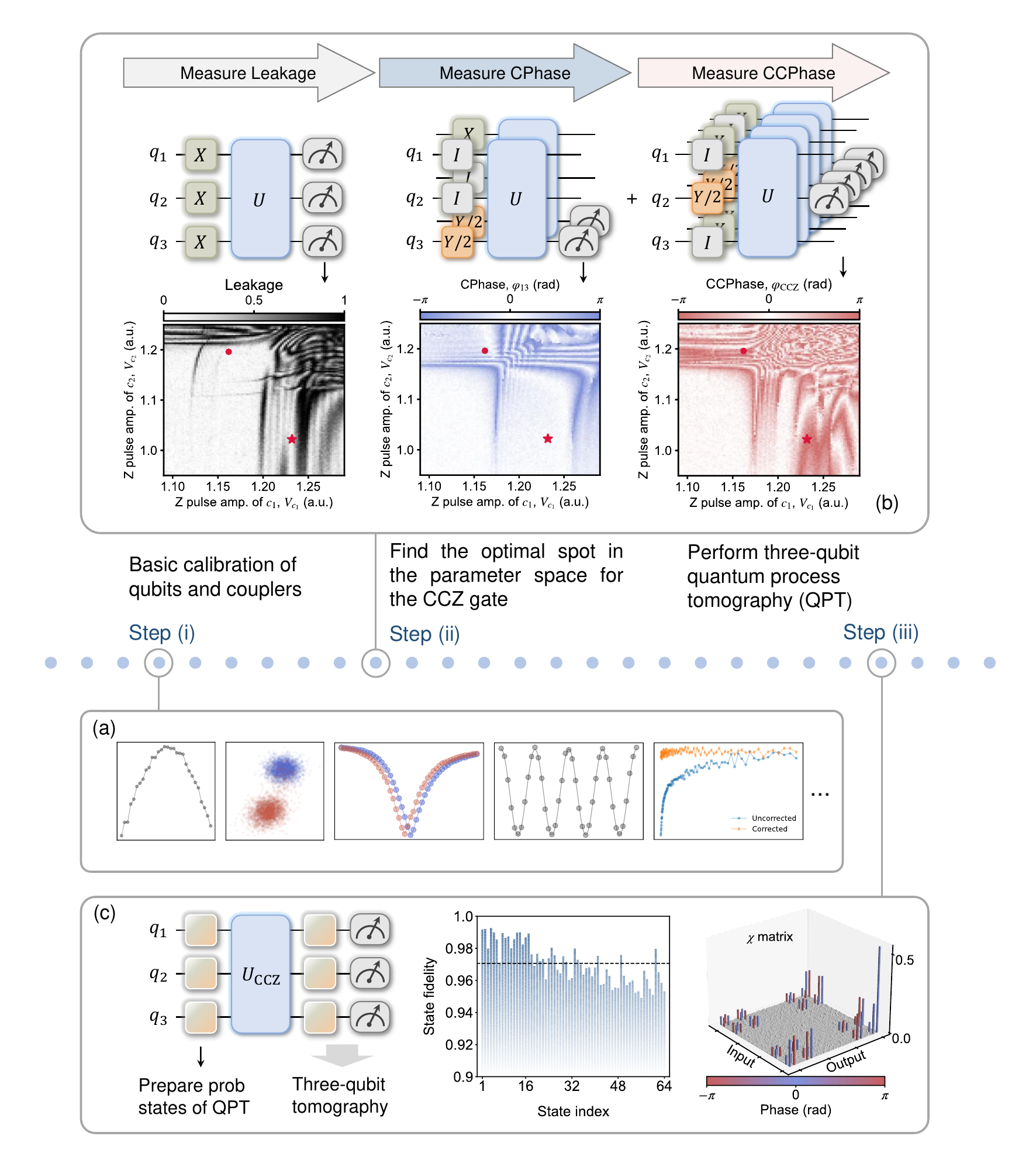}
		\caption{Experimental measurements and calibrations for CCZ gate. (a) Basic calibration of qubits and couplers. (b) Calibration of the Z pulse amplitudes of couplers for the CCZ gate. The optimal operating points, where $\varphi_{\mathrm{CCZ}}=\pm\pi$ while both leakage and $\varphi_{13}$ approach zero, are marked by a dot and a star. Given that the working point marked by the star has a smaller $|\varphi_{13}|$, we select it as the preferred operating point.
		The circuits to measure leakage, CPhase, and CCPhase are illustrated at the middle.
		(c) Experimental data of quantum process tomography (QPT) of the CCZ gate. From left to right are the QPT cirucit, the fidelities of $64$ prob states, and $\chi$-matrix. The average final state fidelity is $97.06\%$ and the process fidelity is $93.54\%$.
%
		 }\label{fig:2}
	\end{center}
\end{figure*}

Our experiment is performed on a 21-qubit flip-chip quantum processor (Fig.~\ref{fig:1}(a)), where every two nearest-neighbor (NN) qubits are coupled through a tunable coupler. For the purposes of our investigation, we select a subset consisting of three qubits ($q_1$, $q_2$, $q_3$) and two interqubit couplers ($c_1$, $c_2$), as schematically shown in Fig.~\ref{fig:1}(c). 
The Hamiltonian of the total system is ($\hbar=1$)
\begin{equation}
    H\!=\!\sum_{i}(\omega_{i}b_{i}^{\dagger}b_{i}\!-\!\frac{\alpha_{i}}{2}b_{i}^{\dagger}b_{i}^{\dagger}b_{i}b_{i})\!-\!\sum_{ij}g_{ij}(b_{i}\!-\!b_{i}^{\dagger})(b_{j}\!-\!b_{j}^{\dagger}),
\end{equation}
where $b_i$ ($i \in \{1,2,3,c_1,c_2\}$) is the annihilation operator and $g_{ij}$ denotes the direct capacitive coupling. Here all qubit frequencies and anharmonicities are fixed, i.e., $\omega_1/2\pi=5.000~\!\mathrm{GHz}$, $\omega_2/2\pi=4.896~\!\mathrm{GHz}$, $\omega_3/2\pi=5.040~\!\mathrm{GHz}$, $\alpha_1/2\pi=-198~\!\mathrm{MHz}$, $\alpha_2/2\pi=-200~\!\mathrm{MHz}$, $\alpha_3/2\pi=-206~\!\mathrm{MHz}$, 
$\alpha_{c_1}/2\pi=-340~\!\mathrm{MHz}$, and $\alpha_{c_2}/2\pi=-320~\!\mathrm{MHz}$. By applying the Z pulses to the couplers, one can dynamically adjust the coupler frequencies to tune the effective couplings between computational qubits. Recent approaches leverage this capability to implement high-fidelity controlled-Z (CZ) gates for superconducting qubits equipped with tunable couplers~\cite{Collodo2020, Xu2020, Xu2021, Ye2021, Ding2023, Barends2019a, sung_realization_2021, Moskalenko2022, Li2025}. 

\begin{figure*}[t]
	\begin{center}
		\includegraphics[width=0.97\textwidth]{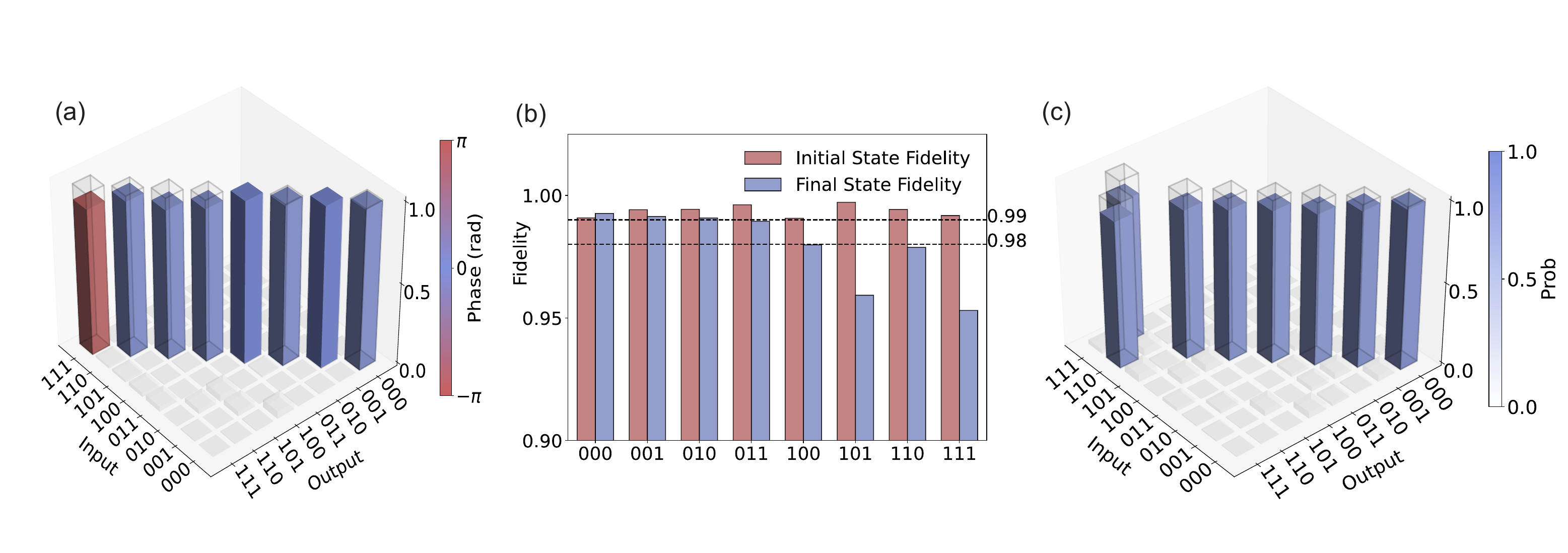}
		\caption{Experimental results for the CCZ and Toffoli gates. 
		(a) The truth table of the CCZ gate. Theoretical probabilities are represented by transparent cylinders, 
		while experimental probabilities and conditional phases are depicted 
		by the height and color of solid cylinders, respectively. The average fidelity of the truth table is $ 96.52\% $.
        (b) The state fidelity of the CCZ gate. The average final state fidelity is $97.94\%$.
		(c) The truth table of the Toffoli gate. Theoretical probabilities appear as hollow cylinders, 
		with experimental probabilities indicated by the color bar. }
		\label{fig:qpt_ccz_ccx}
	\end{center}
\end{figure*}

The routine method for preparing a CCZ gate involves combining a series of single-qubit gates and two-qubit CNOT gates~\cite{m_q_cruz_shallow_2024}, as illustrated in Fig.~\ref{fig:1}(d). In addition to implementing the CCZ gate using this conventional approach, we realize a direct CCZ gate with high scalability, which is composed of two segments (see Fig.~\ref{fig:1}(e)). The first segment $U$ is to achieve the accumulation of the controlled-controlled phase (CCPhase), in which the three qubits are unbiased but two couplers are simultaneously applied Z pulses to generating the three-qubit interaction. Here we set the pulses of these two couplers as flat-top Gaussian waveforms, parametrized by the same duration $\tau$ and the respective maximal amplitudes $V_{c_1}$ and $V_{c_2}$. The calibration of pulse parameters primarily aims to optimize the CCPhase to $\pm\pi$ while minimizing leakage errors (specific calibration details will be described later). When the initial state is set to $\ket{111}$, the calibrated $U$ induces the level repulsion between $\ket{111}$ and all other three-excitation states (e.g., $\ket{102}$ and $\ket{201}$), resulting in a pronounced effective three-body interaction. This engineered $U$ dynamically suppresses leakage from the computational subspace by driving the population back toward $\ket{111}$ (Fig.~\ref{fig:1}(f)), while simultaneously introducing a three-qubit conditional phase.
However, the first segment may also introduce two conditional phases (CPhases) of two pairs of NN qubits, namely $q_1q_2$ and $q_2q_3$. This occurs because changes in the coupler frequency can alter the effective ZZ interaction between the NN qubits connected to the coupler~\cite{Collodo2020,Li2020}, leading to an accumulation of the CPhases over the duration~$\tau$. To compensate for these CPhases accumulated in the first segment, the second segment $U_{\mathrm{CP}}$ involves applying two-qubit CPhase gates sequentially to qubit pairs $q_1q_2$ and $q_2q_3$. The total length of the direct CCZ gate thus becomes $\tau+\tau_{12}+\tau_{23}$, where $\tau_{12}$ and $\tau_{23}$ represent the lengths of the respective CPhase gates.
In the experiment, we take $\tau=150~\!\mathrm{ns}$, $\tau_{12}=62~\!\mathrm{ns}$ and $\tau_{23}=44~\!\mathrm{ns}$, yielding a total length for the direct CCZ gate of $256~\!\mathrm{ns}$. This is notably shorter than the typical total length of $640~\!\mathrm{ns}$ achieved using the routine decomposition strategy. Consequently, the direct CCZ gate is expected to exhibit lower multilayer leakage due to decoherence, as demonstrated in Fig.~\ref{fig:1}(g).

In the following, we outline several key steps for calibrating the direct CCZ gate (see Fig.~\ref{fig:2}). The initial step (\romannumeral1) involves calibrating the individual qubits to ensure that they operate within optimal parameters at idle points. As required for subsequent steps, all single-qubit gates must be calibrated in advance at this stage. Furthermore, the calibration of the coupler Z distortion, including both short-~and~long-time distortion calibration~\cite{Li2025}, is essential for achieving high-fidelity two- and three-qubit gates.
The second step (\romannumeral2) focuses on identifying the optimal spot in the parameter space that yield the best performance for the CCZ gate. This includes optimizing the coupler Z pulse amplitudes $V_{c_1}$ and $V_{c_2}$ for a fixed $\tau=150~\!\mathrm{ns}$ to achieve the desired $\pm\pi$ CCPhase while minimizing leakage errors. To measure the leakage, we prepare $\ket{111}$ by applying $X$ gates to the three qubits and measure the population of $\ket{111}$ after $U$. To efficiently characterize the CCPhase, we initialize three qubits in six special states, i.e., $|0\!+\!0\rangle$,~$|1\!+\!1\rangle$,~$|1\!+\!0\rangle$,~$|0\!+\!1\rangle$,~$|1\,0\ \!+\!\rangle$,~and $|0\,0\ \!+\!\rangle$. We then measure all conditional phases, i.e., $\varphi_{12} = \varphi_{|1+0\rangle} - \varphi_{|0+0\rangle}$, $\varphi_{23} = \varphi_{|0+1\rangle} - \varphi_{|0+0\rangle}$, $\varphi_{13} = \varphi_{|1\,0 \, \!+\!\rangle} - \varphi_{|0\,0\, \!+\!\rangle}$, $\varphi_{123} = \varphi_{|1+1\rangle} - \varphi_{|0+0\rangle}$,
where $\varphi_{12}$, $\varphi_{23}$, and $\varphi_{13}$ represent the two-qubit conditional phases. 
Here $\varphi_{123}$ is the three-qubit conditional phases when both control qubits $q_1$ and $q_3$ are excited to $\ket{1}$. It actually includes the CCPhase of CCZ and all the two-qubit conditional phases. Thus, the CCPhase of CCZ is given by $\varphi_{\mathrm{CCZ}} = \varphi_{123} - \varphi_{13} - \varphi_{12} - \varphi_{23}$.
As mentioned before, $\varphi_{12}$ and $\varphi_{23}$ can be compensated by applying corresponding CPhase gates. However, compensating for $\varphi_{13}$ poses a significant challenge, as it necessitates a nonadjacent CPhase gate between $q_1$ and $q_3$. Although gate decomposition can be employed to achieve this, it may introduce additional complexity and noise, potentially degrading the overall fidelity of CCZ. Therefore, it is crucial to calibrate the system so that $\varphi_{13}$ is as close to zero as possible, and thus $\varphi_{\mathrm{CCZ}}\approx\varphi_{123} - \varphi_{12} - \varphi_{23}$. We try to search for an operating point where $\varphi_{\mathrm{CCZ}}=\pm \pi$ while simultaneously minimizing the leakage and $|\varphi_{13}|$. As shown in Fig.~\ref{fig:2}(b), two relatively symmetrical optimal spots meet these criteria, indicated by the red circle and red star, respectively. Since the optimal spot marked by the star exhibits a smaller $|\varphi_{13}|\approx0.0743$, we select it as our preferred operating point. These tune-up measurements are qualitatively reproduced by time-dependent Hamiltonian simulations for five interacting transmons with 3 levels~\cite{SM}.
Finally, we apply virtual Z gates~\cite{McKay2017_virtual_z} to compensate for the accumulations of single-qubit dynamic phases that accompany the CCZ gate. These dynamic phases can be first roughly characterized through Ramsey experiments on the three qubits, and then numerically optimized using the Nelder-Mead algorithm, with the 
quantum process tomography (QPT) $\chi$-matrix fidelity serving as the objective function. In our QPT experiments, we construct a comprehensive set of $64$ probe states by forming the tensor product of four single-qubit operations $\{I, X, X/2, Y/2\}$ for each of the three qubits ($4^3=64$). The resulting state fidelities and the $\chi$-matrix after the calibration are shown in Fig.~\ref{fig:2}(c), where the average fidelity of these $64$ states after (before) applying a CCZ gate is $97.06\%$ ($99.37\%$), and the process fidelity, calculated as $F_{\chi}=\mathrm{Tr}(\chi_{\mathrm{exp}}\chi_{\mathrm{ideal}})$, is $93.54\%$.

Under the assumption of negligible decoherence and no leakage to the environment, our numerical simulation of the direct CCZ gate, utilizing the time-dependent Hamiltonian, achieves an average state fidelity of $99.45\%$ and a process fidelity of $98.75\%$. These high fidelities underscore the exceptional potential of our proposed scheme. However, the observed discrepancies between the simulation and experimental results are mainly attributed to the influence of decoherence (see Supplemental Material~\cite{SM} for details). As experimental technology advances and the quality of quantum devices improves, it is anticipated that these discrepancies will diminish, further enhancing the performance of the three-qubit gates on large-scale quantum chips.

\begin{figure}[t]
	\begin{center}
		\includegraphics[width=0.41\textwidth]{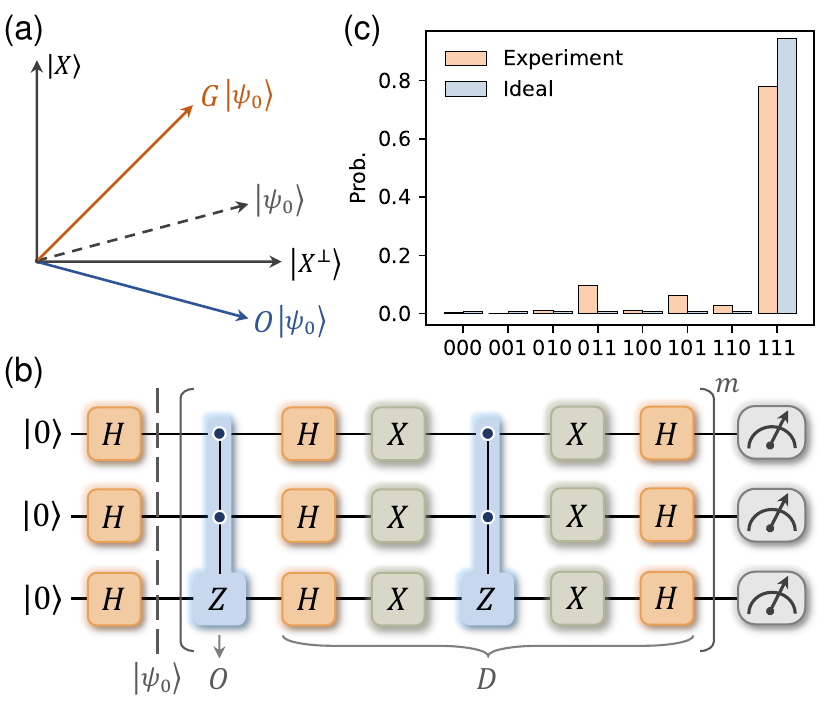}
		\caption{Demonstration of three-qubit Grover search algorithm. 
		(a) Schematic diagram of Grover algorithm principle.
			(b) Quantum circuit of three-qubit Grover search algorithm.
			(c) Probability data after two Grover operations.}
		\label{fig:Grover}
	\end{center}
\end{figure}

To demonstrate the performance of the CCZ gate, we first plot the truth table for computational basis states in Fig.~\ref{fig:qpt_ccz_ccx}(a). The visibility of the measured truth table is $\mathrm{Tr}[U_{\mathrm{exp}}U_{\mathrm{ideal}}]/8=96.52\%$, with an average final state fidelity of $97.94\%$ (Fig.~\ref{fig:qpt_ccz_ccx}(b)), indicating high accuracy and precision in our measurements. Furthermore, by combining the CCZ gate with Hadamard gates, we construct a Toffoli (CCNOT) gate. As shown in Fig.~\ref{fig:qpt_ccz_ccx}(c), the corresponding truth table exhibits a visibility of $92.83\%$, underscoring the versatility and effectiveness of our CCZ gate implementation for advanced multiqubit operations

In addition, we utilize the calibrated CCZ gate to demonstrate an example of the three-qubit Grover search algorithm (GA)~\cite{Grover1997,Figgatt2017}. As a fundamental quantum algorithm, GA leverages quantum coherence as a resource to speed up the process of searching for a target state. It requires the system to be initialized to the maximum superposition state $\ket{\psi_0}$ and repeated by the Grover operator $G=DO$, where $O=1-2\ket{s}\bra{s}$ ($\ket{s}$ is the target state) serves as the oracle and $D=2\ket{\psi_0}\bra{\psi_0}-1$ is diffusion operator, performing an inversion about average operation. The general principle of GA is briefly shown in Fig.~\ref{fig:Grover}(a). In this demonstration, we initialize the qubits with three Hadamard gates to create a superposition of all possible states. We then use the CCZ gate as the oracle to mark the target state $\ket{s}=\ket{111}$. To implement the diffusion operator, we combine the CCZ gate with three Hadamard gates and three $X$ gates, as depicted in Fig.~\ref{fig:Grover}(b).
Theoretically, GA searches for the target state $|111\rangle$ among eight computational states. The optimal number of iterations is given by ${\pi}/{4}\sqrt{{1}/{N}} \approx 2.22$~\cite{nielsen_chuang_2000}, where $N=8$ denotes the search space size. Our experimental results closely align with theoretical predictions, i.e., the probability amplitude of the target state is significantly higher than that of other states after two Grover iterations, as illustrated in Fig.~\ref{fig:Grover}(c). This result demonstrates the effective implementation of the GA using our CCZ gate.

In conclusion, we propose a high-fidelity scheme for implementing a three-qubit CCZ gate in superconducting quantum devices. Our method achieves an average state fidelity of $97.94\%$ and a process fidelity of $93.54\%$, demonstrating its high performance. This method is scalable and requires minimal connectivity between qubits and couplers, offering significant advantages in terms of gate length and leakage compared to the decomposed CCZ gate schemes. We further validate the efficacy of our approach by successfully utilizing the CCZ gate to construct the Toffoli gate and implement the Grover search algorithm. Numerical simulations reveal that our proposed method for implementing the CCZ gate can theoretically attain an average state fidelity of $99.45\%$ and a process fidelity of $98.75\%$, underscoring the outstanding potential of our scheme.
Our work is expected to substantially contribute to the advancement of complex quantum algorithms and the realization of scalable quantum systems by providing a reliable and high-fidelity multiqubit gate operation.


\begin{acknowledgments}
This work was supported by the National Natural Science Foundation of China (Grants Nos. T2121001, 92265207, T2322030, 12122504, 12274142, 92365206, 12104055, 12204528, 92365301, 12404578), the Innovation Program for Quantum Science and Technology (Grant No. 2021ZD0301800), Beijing National Laboratory for Condensed Matter Physics (2024BNLCMPKF022), the Beijing Nova Program (No. 20220484121), and the China Postdoctoral Science Foundation (Grant No. GZB20240815). This work was supported by the Synergetic Extreme Condition User Facility (SECUF). Devices were made at the Nanofabrication Facilities at Institute of Physics, CAS in Beijing.
\end{acknowledgments}
\textit{Data availability}---The data that support the findings of this article are openly available~\cite{shi_2025_15792648}.

%

\end{document}


\title{Supplemental Material for ``Direct Implementation of High-Fidelity Three-Qubit Gates for Superconducting Processor with Tunable Couplers''}
\author{Hao-Tian Liu}
	\thanks{These authors contributed equally to this work.}
    \affiliation{Beijing National Laboratory for Condensed Matter Physics, Institute of Physics, Chinese Academy of Sciences, Beijing 100190, China}
	\affiliation{School of Physical Sciences, University of Chinese Academy of Sciences, Beijing 100049, China}
	
\author{Bing-Jie Chen}
	\thanks{These authors contributed equally to this work.}
    \affiliation{Beijing National Laboratory for Condensed Matter Physics, Institute of Physics, Chinese Academy of Sciences, Beijing 100190, China}
	\affiliation{School of Physical Sciences, University of Chinese Academy of Sciences, Beijing 100049, China}
	
\author{Jia-Chi Zhang}
    \affiliation{Beijing National Laboratory for Condensed Matter Physics, Institute of Physics, Chinese Academy of Sciences, Beijing 100190, China}
	\affiliation{School of Physical Sciences, University of Chinese Academy of Sciences, Beijing 100049, China}
	
\author{Yong-Xi Xiao}
    \affiliation{Beijing National Laboratory for Condensed Matter Physics, Institute of Physics, Chinese Academy of Sciences, Beijing 100190, China}
	\affiliation{School of Physical Sciences, University of Chinese Academy of Sciences, Beijing 100049, China}	
	
\author{Tian-Ming Li}
    \affiliation{Beijing National Laboratory for Condensed Matter Physics, Institute of Physics, Chinese Academy of Sciences, Beijing 100190, China}
	\affiliation{School of Physical Sciences, University of Chinese Academy of Sciences, Beijing 100049, China}		
	
\author{Kaixuan Huang}
    \affiliation{Beijing Key Laboratory of Fault-Tolerant Quantum Computing, Beijing Academy of Quantum Information Sciences, Beijing 100193, China}

\author{Ziting Wang}
    \affiliation{Beijing Key Laboratory of Fault-Tolerant Quantum Computing, Beijing Academy of Quantum Information Sciences, Beijing 100193, China}

\author{Hao Li}
    \affiliation{Beijing Key Laboratory of Fault-Tolerant Quantum Computing, Beijing Academy of Quantum Information Sciences, Beijing 100193, China}
    
\author{Kui Zhao}
    \affiliation{Beijing Key Laboratory of Fault-Tolerant Quantum Computing, Beijing Academy of Quantum Information Sciences, Beijing 100193, China}
    
\author{Yueshan Xu}
    \affiliation{Beijing Key Laboratory of Fault-Tolerant Quantum Computing, Beijing Academy of Quantum Information Sciences, Beijing 100193, China}
    
\author{Cheng-Lin Deng}
    \affiliation{Beijing National Laboratory for Condensed Matter Physics, Institute of Physics, Chinese Academy of Sciences, Beijing 100190, China}
	\affiliation{School of Physical Sciences, University of Chinese Academy of Sciences, Beijing 100049, China}	
	
\author{Gui-Han Liang}
    \affiliation{Beijing National Laboratory for Condensed Matter Physics, Institute of Physics, Chinese Academy of Sciences, Beijing 100190, China}
	\affiliation{School of Physical Sciences, University of Chinese Academy of Sciences, Beijing 100049, China}		
	
\author{Zheng-He Liu}
    \affiliation{Beijing National Laboratory for Condensed Matter Physics, Institute of Physics, Chinese Academy of Sciences, Beijing 100190, China}
	\affiliation{School of Physical Sciences, University of Chinese Academy of Sciences, Beijing 100049, China}		
	
\author{Si-Yun Zhou}
    \affiliation{Beijing National Laboratory for Condensed Matter Physics, Institute of Physics, Chinese Academy of Sciences, Beijing 100190, China}
	\affiliation{School of Physical Sciences, University of Chinese Academy of Sciences, Beijing 100049, China}		
	
\author{Cai-Ping Fang}
    \affiliation{Beijing National Laboratory for Condensed Matter Physics, Institute of Physics, Chinese Academy of Sciences, Beijing 100190, China}
	\affiliation{School of Physical Sciences, University of Chinese Academy of Sciences, Beijing 100049, China}		
	
\author{Xiaohui Song}
    \affiliation{Beijing National Laboratory for Condensed Matter Physics, Institute of Physics, Chinese Academy of Sciences, Beijing 100190, China}
    \affiliation{School of Physical Sciences, University of Chinese Academy of Sciences, Beijing 100049, China}
    \affiliation{Hefei National Laboratory, Hefei 230088, China}
    
\author{Zhongcheng Xiang}
    \affiliation{Beijing National Laboratory for Condensed Matter Physics, Institute of Physics, Chinese Academy of Sciences, Beijing 100190, China}
    \affiliation{School of Physical Sciences, University of Chinese Academy of Sciences, Beijing 100049, China}	
    \affiliation{Hefei National Laboratory, Hefei 230088, China}
    
\author{Dongning Zheng}
    \affiliation{Beijing National Laboratory for Condensed Matter Physics, Institute of Physics, Chinese Academy of Sciences, Beijing 100190, China}
    \affiliation{School of Physical Sciences, University of Chinese Academy of Sciences, Beijing 100049, China}
    \affiliation{Hefei National Laboratory, Hefei 230088, China}

\author{Yun-Hao Shi}
    \thanks{yhshi@iphy.ac.cn}
    \affiliation{Beijing National Laboratory for Condensed Matter Physics, Institute of Physics, Chinese Academy of Sciences, Beijing 100190, China}
    \affiliation{Hefei National Laboratory, Hefei 230088, China}

\author{Kai Xu}
    \thanks{kaixu@iphy.ac.cn}
    \affiliation{Beijing National Laboratory for Condensed Matter Physics, Institute of Physics, Chinese Academy of Sciences, Beijing 100190, China}
    \affiliation{School of Physical Sciences, University of Chinese Academy of Sciences, Beijing 100049, China}
    \affiliation{Beijing Key Laboratory of Fault-Tolerant Quantum Computing, Beijing Academy of Quantum Information Sciences, Beijing 100193, China}
    \affiliation{Hefei National Laboratory, Hefei 230088, China}
    \affiliation{Songshan Lake Materials Laboratory, Dongguan 523808, Guangdong, China}

\author{Heng Fan}
    \thanks{hfan@iphy.ac.cn}
    \affiliation{Beijing National Laboratory for Condensed Matter Physics, Institute of Physics, Chinese Academy of Sciences, Beijing 100190, China}
    \affiliation{School of Physical Sciences, University of Chinese Academy of Sciences, Beijing 100049, China}
    \affiliation{Beijing Key Laboratory of Fault-Tolerant Quantum Computing, Beijing Academy of Quantum Information Sciences, Beijing 100193, China}
    \affiliation{Hefei National Laboratory, Hefei 230088, China}
    \affiliation{Songshan Lake Materials Laboratory, Dongguan 523808, Guangdong, China}
\maketitle
\beginsupplement
\tableofcontents
\newpage

\section{\label{sec:setup} Experimental setup}
\begin{figure}[ht!]
	\includegraphics[width=1.01\textwidth]{ 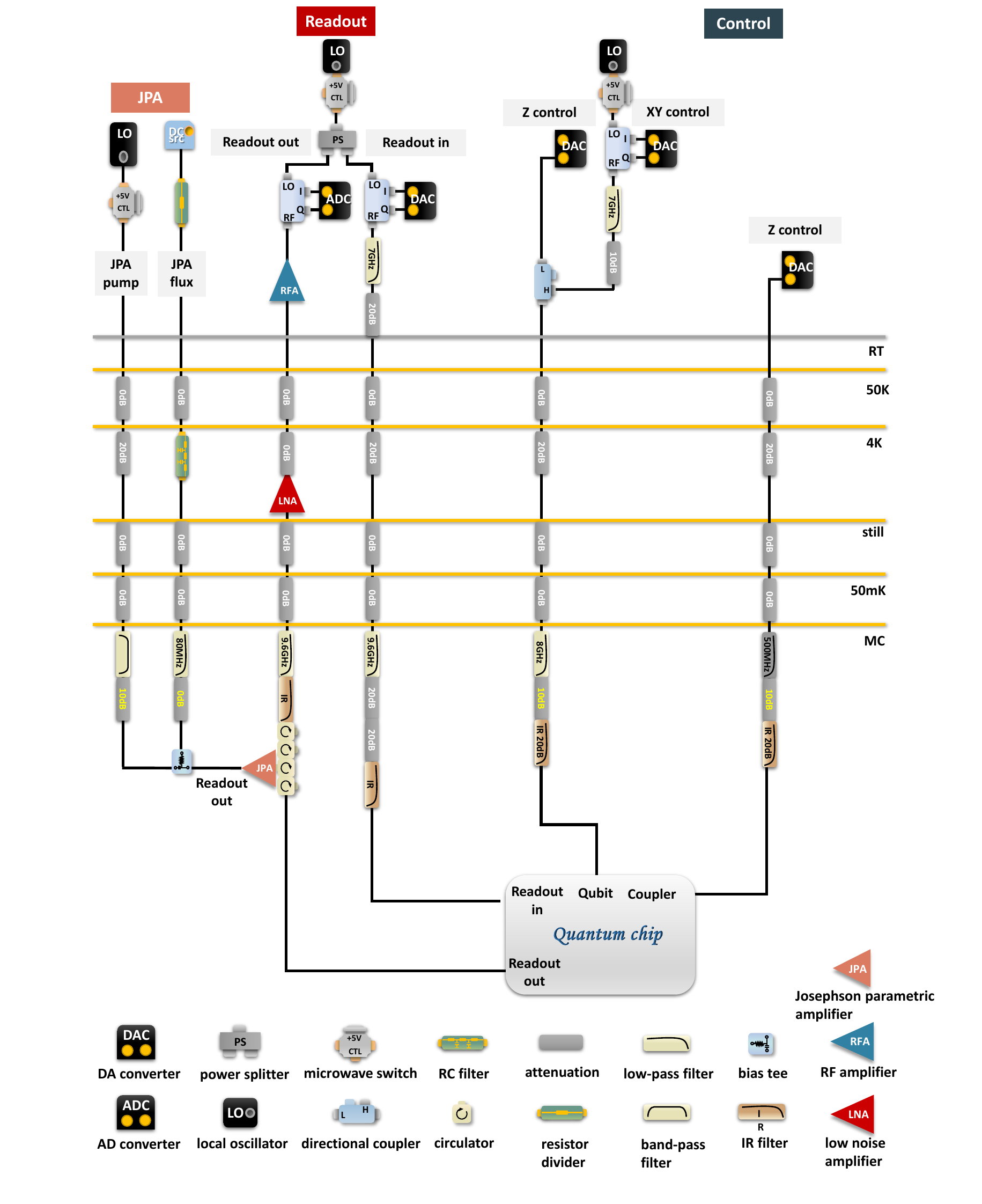}
	\caption{A schematic diagram of the experimental system and wiring information. }\label{fig:fig0}
\end{figure}

Our 21-qubits superconducting quantum chip is operated within a BlueFors dilution refrigerator, with a mixing chamber maintained at approximately $20~\!\mathrm{mK}$. The typical configuration of the control electronics and cryogenic equipment is illustrated in Fig.~\ref{fig:fig0}. The system includes two readout transmission lines, each equipped with a superconducting Josephson parametric amplifier (JPA) at $20~\!\mathrm{mK}$, a cryogenic low-noise amplifier (LNA) at $4~\!\mathrm{K}$, and a room-temperature RF amplifier (RFA). The readout pulse is generated by a arbitrary waveform generator (AWG), which consists of two digital-to-analog converter (DAC) channels and a local oscillator (LO). To mitigate thermal photons from higher-temperature stages, the pulse is filtered using RF attenuators at different temperatures, and finally captured by the analog-to-digital converter (ADC).

To optimize space in the refrigerator, we use directional couplers to combine the high-frequency XY signal with the low-frequency Z bias signal at room temperature, which enables us to merge two control lines of the qubits into a single line at cryogenic temperatures. Additionally, microwave switches are placed at each LO port to mitigate thermal excitations caused by constant microwave signal.

The qubit parameters are summarized in Table \ref{tab:paras}. We optimize the frequency arrangement by considering the two-level system (TLS) and residual coupling between qubits, in order to achieve a long coherence time and minimize XY crosstalk.
In our experiment, the direct qubit-coupler hopping coupling is $g/2\pi \approx90~\!\mathrm{MHz}$, whereas the direct hopping coupling between nearest-neighbor (NN) qubits is $g/2\pi\approx 5~\!\mathrm{MHz}$. The idle frequencies of couplers are optimized to a point where the effective static qubit-qubit coupling is nearly zero.

\begin{table}[htbp!]
	\begin{ruledtabular}
		\begin{tabular}{cccc}
			Qubit & {$Q_1$} & {$Q_2$} & $Q_3$ \\[.02cm] \hline
	        Qubit Frequency, $f_{10}$($\mathrm{GHz}$)& 5.000& 4.896& 5.040\\
	        Readout Frequency,  $f_{r}$($\mathrm{GHz}$)& 6.787& 6.862& 6.830\\
	        Anharmonicity, $\alpha$ ($\mathrm{MHz}$)& -198& -200& -206\\
	        Relaxation Time, $T_1$ ($\mathrm{\mu s}$)& 37.76& 43.34& 59.13\\
	        Readout Fidelity of $|0\rangle$, $F_0$& 98.14\%& 97.68\%& 98.61\%\\
	        Readout Fidelity of $|1\rangle$, $F_1$& 95.03\%& 90.10\%& 92.17\%\\
		\end{tabular}
	\end{ruledtabular}
	\caption{Device parameters for qubits.}\label{tab:paras}
\end{table}

\section{\label{sec:ham} The effective Hamiltonian}
\begin{figure}[t]
	\includegraphics[width=0.6\textwidth]{ 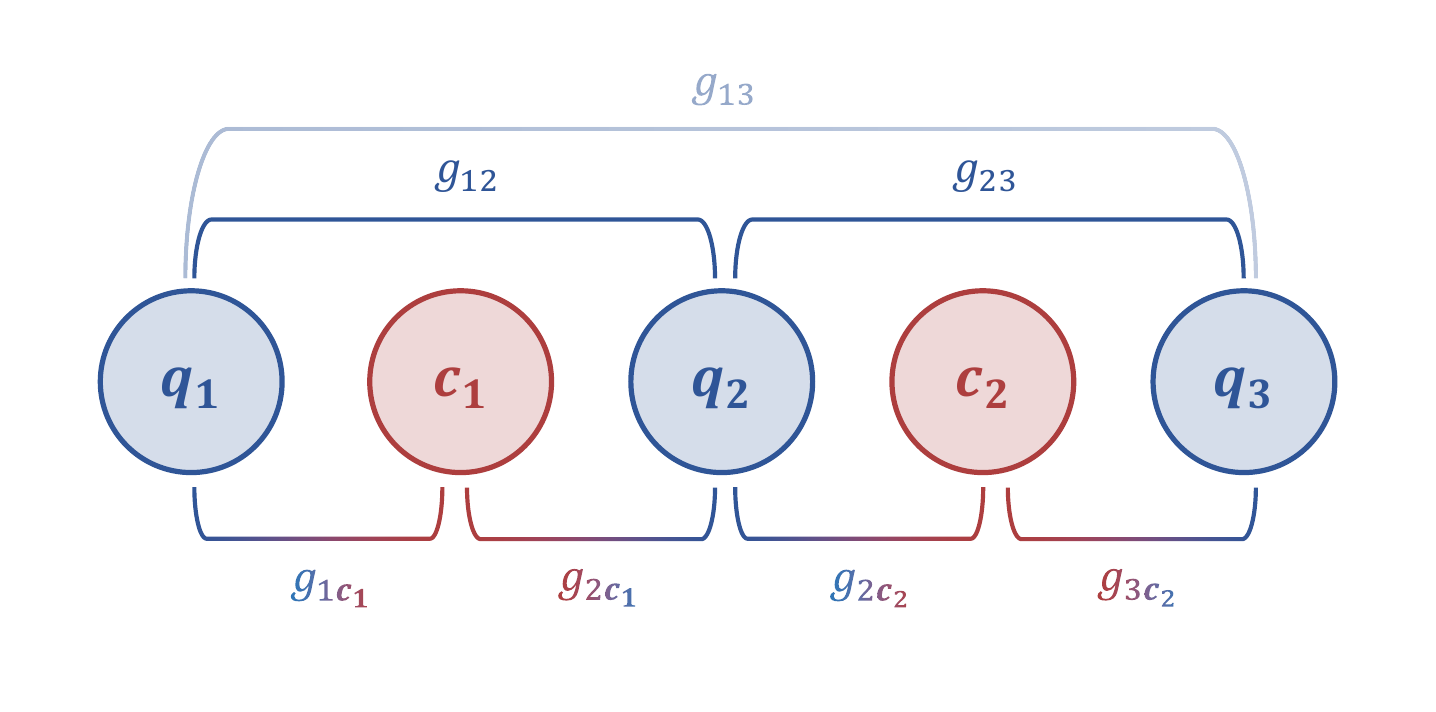}
	\caption{Sketch of a five-body system in a superconducting quantum chip. }\label{fig:qcqcq}
\end{figure}

Our system can be described by five interacting transmons (three qubits and two couplers, see Fig.~\ref{fig:qcqcq}):
\begin{eqnarray}
H&=&H_{q}+H_{c}+V_{qq}+V_{qc},\\
H_{q}&=&\omega_{1}b_{1}^{\dagger}b_{1}+\frac{\alpha_{1}}{2}b_{1}^{\dagger}b_{1}^{\dagger}b_{1}b_{1}+\omega_{2}b_{2}^{\dagger}b_{2}+ {\frac{\alpha_{2}}{2}}b_{2}^{\dagger}b_{2}^{\dagger}b_{2}b_{2}+\omega_{3}{b_{3}^{\dagger}}b_{3}+{\frac{\alpha_{3}}{2}}b_{3}^{\dagger}b_{3}^{\dagger}b_{3}b_{3},\\
H_{c}&=&\omega_{c_1}b_{c_1}^{\dagger}b_{c_1}+\frac{\alpha_{c_1}}{2}b_{c_1}^{\dagger}b_{c_1}^{\dagger}b_{c_1}b_{c_1}+\omega_{c_2}b_{c_2}^{\dagger}b_{c_2}+\frac{\alpha_{c_2}}{2}b_{c_2}^{\dagger}b_{c_2}^{\dagger}b_{c_2}b_{c_2},\\
V_{qq}&=&g_{12}(b_{1}^{\dagger}-b_{1})(b_{2}-b_{2}^{\dagger})+g_{23}(b_{2}^{\dagger}-b_{2})(b_{3}-b_{3}^{\dagger})+g_{13}(b_{1}^{\dagger}-b_{1})(b_{3}-b_{3}^{\dagger}),\\
V_{qc}&=&g_{1c_1}(b_{1}^{\dagger}-b_{1})(b_{c_1}-b_{c_1}^{\dagger})+g_{2c_1}(b_{2}^{\dagger}-b_{2})(b_{c_1}-b_{c_1}^{\dagger})+g_{2c_2}(b_{2}^{\dagger}-b_{2})(b_{c_2}-b_{c_2}^{\dagger})+g_{3c_2}(b_{3}^{\dagger}-b_{3})(b_{c_2}-b_{c_2}^{\dagger}),
\end{eqnarray}
where $b_i$ is the annihilation operator, $\omega_i$ is the qubit (or coupler) frequency, $\alpha_i$ is the anharmonicity, and $g_{ij}$ denotes the direct qubit-qubit (or qubit-coupler) hopping coupling. To decouple the couplers from the system, we apply the Schrieffer-Wolff transformation and consider the high-order perturbation contributions. This approach allows us to derive an effective Hamiltonian that focuses on the interactions between the three computational qubits, i.e.,
\begin{eqnarray}
V^{\mathrm{eff}}_{qq}&=&V^{\mathrm{XY}}_{qq}+V^{\mathrm{ZZ}}_{qq},\\
V^{\mathrm{XY}}_{qq}&=&\tilde{g}_{12}(b_{1}^{\dagger}b_{2}+b_{1}b_{2}^{\dagger})+\tilde{g}_{23}(b_{2}^{\dagger}b_{3}+b_{2}b_{3}^{\dagger})+\tilde{g}_{13}(b_{1}^{\dagger}b_{3}+b_{1}b_{3}^{\dagger}),\\
V^{\mathrm{ZZ}}_{qq}&=&\zeta_{12}b_{1}^{\dagger}b_{1}b_{2}^{\dagger}b_{2}+\zeta_{23}b_{2}^{\dagger}b_{2}b_{3}^{\dagger}b_{3}+\zeta_{13}b_{1}^{\dagger}b_{1}b_{3}^{\dagger}b_{3}+\zeta_{123}b_{1}^{\dagger}b_{1}b_{2}^{\dagger}b_{2}b_{3}^{\dagger}b_{3},
\end{eqnarray}
where $\tilde{g}_{ij}$ denotes the effective XY (hopping) coupling, $\zeta_{ij}$ signifies the parasitic two-body ZZ coupling, and $\zeta_{123}$ represents the three-body ZZZ coupling. In the next section, we will demonstrate detailed derivation of using perturbation theory to calculate these diagonal interaction terms. Note that these effetive couplings depend on the detuning between the qubits and their intermediate coupler, which means we can adjust the coupler frequency to tune the effective couplings between the computational qubits. However, when qubits are detuned, they can barely exchange particles via the effective hopping. Thus, we can neglect $V^{\mathrm{XY}}_{qq}$ and
describe the system dynamics govered by $V^{\mathrm{ZZ}}_{qq}$. Considering only two levels (computational subspace), the evolution of the system over a duration $\tau$ corresponds to
\begin{equation}\label{eq:U}
	U=e^{-{\mathrm{i}}\int_{0}^{\tau}V^{\mathrm{ZZ}}_{qq}\mathrm{d}t}=
	\begin{pmatrix}
		1 & ~ & ~ & ~ & ~ & ~ & ~ & ~\\
		~ & 1 & ~ & ~ & ~ & ~ & ~ & ~\\
		~ & ~ & 1 & ~ & ~ & ~ & ~ & ~\\
		~ & ~ & ~ & e^{-\mathrm{i}\varphi_{23}} & ~ & ~ & ~ & ~\\
		~ & ~ & ~ & ~ & 1 & ~ & ~ & ~\\
		~ & ~ & ~ & ~ & ~ & e^{-\mathrm{i}\varphi_{13}} & ~ & ~\\
		~ & ~ & ~ & ~ & ~ & ~ & e^{-\mathrm{i}\varphi_{12}} & ~\\
		~ & ~ & ~ & ~ & ~ & ~ & ~ & e^{-\mathrm{i}\varphi_{123}}
	\end{pmatrix},
\end{equation}
with $\varphi_{123}=\varphi_{12}+\varphi_{23}+\varphi_{13}+\varphi_{\mathrm{CCZ}}$, where $\varphi_{ij}=\int_{0}^{\tau}\zeta_{ij}\mathrm{d}t$ and $\varphi_{\mathrm{CCZ}}=\int_{0}^{\tau}\zeta_{123}\mathrm{d}t$ represent two-qubit and three-qubit conditional phases, repsectively. Eq.~\eqref{eq:U} actually describes the first segment of our CCZ gate in the absence of decoherence and leakage. The second segment consists of two CPhase gates to compensate for $\varphi_{12}$ and $\varphi_{23}$, namely

\begin{equation}\label{eq:U_CP}
	U_{\mathrm{CP}}=e^{\mathrm{i}\varphi_{23}b_{2}^{\dagger}b_{2}b_{3}^{\dagger}b_{3}}e^{\mathrm{i}\varphi_{12}b_{1}^{\dagger}b_{1}b_{2}^{\dagger}b_{2}}=
	\begin{pmatrix}
		1 & ~ & ~ & ~ & ~ & ~ & ~ & ~\\
		~ & 1 & ~ & ~ & ~ & ~ & ~ & ~\\
		~ & ~ & 1 & ~ & ~ & ~ & ~ & ~\\
		~ & ~ & ~ & e^{\mathrm{i}\varphi_{23}} & ~ & ~ & ~ & ~\\
		~ & ~ & ~ & ~ & 1 & ~ & ~ & ~\\
		~ & ~ & ~ & ~ & ~ & 1 & ~ & ~\\
		~ & ~ & ~ & ~ & ~ & ~ & e^{\mathrm{i}\varphi_{12}} & ~\\
		~ & ~ & ~ & ~ & ~ & ~ & ~ & {e^{\mathrm{i}(\varphi_{12}+\varphi_{23})}}
	\end{pmatrix}.
\end{equation}
If we calibrate the system such that the conditions ${\varphi_{\mathrm{CCZ}}}+\varphi_{13}=\pm\pi$ and $\varphi_{13}\rightarrow{0}$ are satisfied, we can effectively implement a CCZ gate:
\begin{equation}\label{eq:U_CP}
	U_{\mathrm{CCZ}}=U_{\mathrm{CP}}U=	\begin{pmatrix}
		1 & ~ & ~ & ~ & ~ & ~ & ~ & ~\\
		~ & 1 & ~ & ~ & ~ & ~ & ~ & ~\\
		~ & ~ & 1 & ~ & ~ & ~ & ~ & ~\\
		~ & ~ & ~ & 1 & ~ & ~ & ~ & ~\\
		~ & ~ & ~ & ~ & 1 & ~ & ~ & ~\\
		~ & ~ & ~ & ~ & ~ & e^{-\mathrm{i}\varphi_{13}} & ~ & ~\\
		~ & ~ & ~ & ~ & ~ & ~ & 1 & ~\\
		~ & ~ & ~ & ~ & ~ & ~ & ~ & {e^{-\mathrm{i}(\varphi_{\mathrm{CCZ}}+\varphi_{13})}}
	\end{pmatrix}\simeq
	\begin{pmatrix}
		1 & ~ & ~ & ~ & ~ & ~ & ~ & ~\\
		~ & 1 & ~ & ~ & ~ & ~ & ~ & ~\\
		~ & ~ & 1 & ~ & ~ & ~ & ~ & ~\\
		~ & ~ & ~ & 1 & ~ & ~ & ~ & ~\\
		~ & ~ & ~ & ~ & 1 & ~ & ~ & ~\\
		~ & ~ & ~ & ~ & ~ & 1 & ~ & ~\\
		~ & ~ & ~ & ~ & ~ & ~ & 1 & ~\\
		~ & ~ & ~ & ~ & ~ & ~ & ~ & -1
	\end{pmatrix}.
\end{equation}

\section{\label{sec:zz} Perturbations of the diagonal interaction terms}
To further illustrate the principle of the CCZ gate, we analyze the effective two-body ZZ and three-body ZZZ couplings using perturbation theory, i.e., $\zeta=\zeta^{(2)}+\zeta^{(3)}+\zeta^{(4)}+\cdots$, where $\zeta^{(n)}$ represents the contribution at the $n$-th order of perturbation~\cite{PhysRevApplied.16.024037}. In this discussion, we apply the rotating-wave approximation to eliminate fast-oscillating terms, thereby simplifying the system Hamiltonian to
\begin{align}
	H_0&=H_q+H_c=\sum_{j} \left( \omega_j b_j^{\dagger}b_j+ \frac{\alpha_j}{2} b_j^{\dagger} b_j^{\dagger} b_j b_j \right),\\
	V&=V_{qq}+V_{qc}=\sum_{j\neq k} g_{jk} \left( b_j^{\dagger}b_k + b_j b_k^{\dagger}  \right),
\end{align}
where $j,k=1,2,3,c_1,c_2$. The $n$-th order perturbations can be defined as ~\cite{baker_single_2022}
\begin{align}
	&\zeta_{12}^{(n)}=E_{|110\rangle}^{(n)}-E_{|100\rangle}^{(n)}-E_{|010\rangle}^{(n)}+E_{|000\rangle}^{(n)},\\
	&\zeta_{23}^{(n)}=E_{|011\rangle}^{(n)}-E_{|010\rangle}^{(n)}-E_{|001\rangle}^{(n)}+E_{|000\rangle}^{(n)},\\
	&\zeta_{13}^{(n)}=E_{|101\rangle}^{(n)}-E_{|100\rangle}^{(n)}-E_{|001\rangle}^{(n)}+E_{|000\rangle}^{(n)},\\
	&\zeta_{123}^{(n)}=E_{|111\rangle}^{(n)}-E_{|100\rangle}^{(n)}-E_{|010\rangle}^{(n)}-E_{|001\rangle}^{(n)}+2 E_{|000\rangle}^{(n)}.
\end{align}
For the second order, we find 
\begin{align}
	&\zeta_{12}^{(2)} = 2 g_{12}^2 \left( \frac{1}{\Delta_{12} - \alpha_2} - \frac{1}{\Delta_{12}+\alpha_1} \right),\\
	&\zeta_{23}^{(2)} = 2 g_{23}^2 \left( \frac{1}{\Delta_{23} - \alpha_3} - \frac{1}{\Delta_{23}+\alpha_2} \right),\\
	&\zeta_{13}^{(2)} = 2 g_{13}^2 \left( \frac{1}{\Delta_{13} - \alpha_3} - \frac{1}{\Delta_{13}+\alpha_1} \right),\\
	&\zeta_{123}^{(2)} = \zeta_{12}^{(2)} + \zeta_{23}^{(2)} + \zeta_{13}^{(2)},
\end{align}
where $\Delta_{ij}=\omega_{i}-\omega_j$. For the third order, the results are
\begin{align}
	\zeta_{12}^{(3)} &= 4 g_{12} g_{13} g_{23} \left(\frac{1}{\Delta_{13}(\Delta_{12}-\alpha_2)} -\frac{1}{\Delta_{23}( \Delta_{12}+\alpha_1 )}\right) + \frac{2 g_{12} g_{13} g_{23}}{\Delta_{13}  \Delta_{23}} \notag\\
	&\quad + 4 g_{12} g_{1c_1} g_{2c_1} \left(  \frac{1}{\Delta_{1 c_1}(\Delta_{12} -\alpha_2 )} -\frac{1}{\Delta_{2 c_1}(\Delta_{12}+\alpha_1)}+\frac{1}{\Delta_{1 c_1}\Delta_{2 c_1}}\right),
\end{align}	
\begin{align}
	\zeta_{23}^{(3)} &= 4 g_{12} g_{13} g_{23} \left( \frac{1}{\Delta_{13}(\Delta_{23}+\alpha_2)} - \frac{1}{\Delta_{12}(\Delta_{23} - \alpha_3)} \right)+\frac {2 g_{12} g_{13} g_{23}}{\Delta_{12} \Delta_{13}}  \notag\\
	&\quad + 4 g_{23} g_{2c_2} g_{3c_2} \left( \frac{1}{\Delta_{2 c_2}(\Delta_{23} - \alpha_3)}-\frac{1}{\Delta_{3 c_2}(\Delta_{23}+\alpha_2 )} +\frac{1}{\Delta_{2 c_2}\Delta_{3 c_2}}\right),
\end{align}
\begin{align}
	\zeta_{13}^{(3)} &= 4 g_{12} g_{13} g_{23} \left( \frac{1}{\Delta_{23}(\Delta_{13}+\alpha_1)} + \frac{1}{\Delta_{12}(\Delta_{13} - \alpha_3)} - \frac{1}{\Delta_{12}\Delta_{23}}\right),
\end{align}
\begin{align}
	\zeta_{123}^{(3)} &= 4 g_{12} g_{13} g_{23} \left[ \frac{1}{(\Delta_{12}+\alpha_1)(\Delta_{13}+\alpha_1)}+\frac{1}{(\Delta_{13} - \alpha_3)(\Delta_{23} - \alpha_3)} - \frac{1}{(\Delta_{12} - \alpha_2)(\Delta_{23}+\alpha_2 )} \right. \notag\\
	&\quad \left. + \frac{2}{(\Delta_{13}+\alpha_1)(\Delta_{23}+\alpha_2 )} + \frac{2}{(\Delta_{12} - \alpha_2)(\Delta_{13} - \alpha_3)}- \frac{2}{(\Delta_{12}+\alpha_1)(\Delta_{23} - \alpha_3)} \right] \notag\\
	&\quad + 4 g_{12} g_{1c_1} g_{2c_1} \left(\frac{1}{\Delta_{1 c_1}(\Delta_{12}-\alpha_2)} -\frac{1}{\Delta_{2 c_1}(\Delta_{12}+\alpha_1)}+ \frac{1}{\Delta_{1 c_1}\Delta_{2 c_1}}\right) \notag\\
	&\quad + 4 g_{23} g_{2c_2} g_{3c_2} \left( \frac{1}{\Delta_{2 c_2}(\Delta_{23} - \alpha_3)}-\frac{1}{\Delta_{3 c_2}(\Delta_{23}+\alpha_2 )} +\frac{1}{\Delta_{2 c_2}\Delta_{3 c_2}}\right).
\end{align}


Considering the fourth order, the corresponding perturbation terms are expressed as
\begin{align}
	\zeta_{12}^{(4)} &= g_{1c_1}^2 g_{2c_1}^2 \left[  \left( \frac{1}{\Delta_{1c_1}} + \frac{1}{\Delta_{2c_1}} \right)^2 \frac{2}{\Delta_{1c_1} + \Delta_{2c_1} - \alpha_{c_1}}  + \frac{2}{\Delta_{1c_1}^2 (\Delta_{12} - \alpha_2)} - \frac{2}{\Delta_{2c_1}^2 (\Delta_{12} + \alpha_1)}\right. \notag\\
	& \left. \quad  + \left( \frac{1}{\Delta_{2c_1}} - \frac{1}{\Delta_{12}} \right) \frac{1}{\Delta_{1c_1}^2} + \left( \frac{1}{\Delta_{1c_1}} + \frac{1}{\Delta_{12}} \right) \frac{1}{\Delta_{2c_1}^2} \right]\notag\\
	& \quad + g_{1c_1}^2 g_{2c_2}^2 \left[ \frac{1}{\Delta_{1c_1} + \Delta_{2c_2} }\left( \frac{1}{\Delta_{1c_1}} +\frac{1}{\Delta_{2c_2}} \right)^2  + \frac{1}{\Delta_{1c_1}\Delta_{2c_2}^2} + \frac{1}{\Delta_{1c_1}^2\Delta_{2c_2}} \right],
\end{align}
\begin{align}
\zeta_{23}^{(4)}=&g_{2c_2}^2 g_{3c_2}^2 \left[  \left( \frac{1}{\Delta_{2c_2}} + \frac{1}{\Delta_{3c_2}} \right)^2 \frac{2}{\Delta_{2c_2} + \Delta_{3c_2} - \alpha_{c_2}}  + \frac{2}{\Delta_{2c_2}^2 (\Delta_{23} - \alpha_3)} - \frac{2}{\Delta_{3c_2}^2 (\Delta_{23} + \alpha_3)}\right. \notag\\
& \left. \quad  + \left( \frac{1}{\Delta_{3c_2}} - \frac{1}{\Delta_{23}} \right) \frac{1}{\Delta_{2c_2}^2} + \left( \frac{1}{\Delta_{2c_2}} + \frac{1}{\Delta_{23}} \right) \frac{1}{\Delta_{3c_2}^2} \right]\notag\\
& \quad + g_{2c_1}^2 g_{3c_2}^2 \left[ \frac{1}{\Delta_{2c_1} + \Delta_{3c_2} }\left( \frac{1}{\Delta_{2c_1}} +\frac{1}{\Delta_{3c_2}} \right)^2  + \frac{1}{\Delta_{2c_1}\Delta_{3c_2}^2} + \frac{1}{\Delta_{2c_1}^2\Delta_{3c_2}} \right],
\end{align}
\begin{align}
\zeta_{13}^{(4)}=& g_{1c_1}^2 g_{3c_2}^2 \left[ \frac{1}{\Delta_{1c_1} + \Delta_{3c_2} }\left( \frac{1}{\Delta_{1c_1}} +\frac{1}{\Delta_{3c_2}} \right)^2  + \frac{1}{\Delta_{1c_1}\Delta_{3c_2}^2} + \frac{1}{\Delta_{1c_1}^2\Delta_{3c_2}} \right],
\end{align}
\begin{align}
	\zeta_{123}^{(4)}=& g_{1c_1}^2 g_{2c_1}^2 \left[  \left( \frac{1}{\Delta_{1c_1}} + \frac{1}{\Delta_{2c_1}} \right)^2 \frac{2}{\Delta_{1c_1} + \Delta_{2c_1} - \alpha_{c_1}}  + \frac{2}{\Delta_{1c_1}^2 (\Delta_{12} - \alpha_2)} - \frac{2}{\Delta_{2c_1}^2 (\Delta_{12} + \alpha_1)}\right. \notag\\
	& \left. \quad  + \left( \frac{1}{\Delta_{2c_1}} - \frac{1}{\Delta_{12}} \right) \frac{1}{\Delta_{1c_1}^2} + \left( \frac{1}{\Delta_{1c_1}} + \frac{1}{\Delta_{12}} \right) \frac{1}{\Delta_{2c_1}^2} \right]\notag\\
	& \quad + g_{1c_1}^2 g_{2c_2}^2 \left[ \frac{1}{\Delta_{1c_1} + \Delta_{2c_2} }\left( \frac{1}{\Delta_{1c_1}} +\frac{1}{\Delta_{2c_2}} \right)^2  + \frac{1}{\Delta_{1c_1}\Delta_{2c_2}^2} + \frac{1}{\Delta_{1c_1}^2\Delta_{2c_2}} \right]\notag\\
	&\quad+g_{2c_2}^2 g_{3c_2}^2 \left[ 2 \left( \frac{1}{\Delta_{2c_2}} + \frac{1}{\Delta_{3c_2}} \right)^2 \frac{1}{\Delta_{2c_2} + \Delta_{3c_2} - \alpha_{c_2}}  + \frac{2}{\Delta_{2c_2}^2 (\Delta_{23} - \alpha_3)} - \frac{2}{\Delta_{3c_2}^2 (\Delta_{23} + \alpha_3)}\right. \notag\\
	& \left. \quad  + \left( \frac{1}{\Delta_{3c_2}} - \frac{1}{\Delta_{23}} \right) \frac{1}{\Delta_{2c_2}^2} + \left( \frac{1}{\Delta_{2c_2}} + \frac{1}{\Delta_{23}} \right) \frac{1}{\Delta_{3c_2}^2} \right]\notag\\
	& \quad + g_{2c_1}^2 g_{3c_2}^2 \left[ \frac{1}{\Delta_{2c_1} + \Delta_{3c_2} }\left( \frac{1}{\Delta_{2c_1}} +\frac{1}{\Delta_{3c_2}} \right)^2  + \frac{1}{\Delta_{2c_1}\Delta_{3c_2}^2} + \frac{1}{\Delta_{2c_1}^2\Delta_{3c_2}} \right]\notag\\
	&\quad+ g_{1c_1}^2 g_{3c_2}^2 \left[ \frac{1}{\Delta_{1c_1} + \Delta_{3c_2} }\left( \frac{1}{\Delta_{1c_1}} +\frac{1}{\Delta_{3c_2}} \right)^2  + \frac{1}{\Delta_{1c_1}\Delta_{3c_2}^2} + \frac{1}{\Delta_{1c_1}^2\Delta_{3c_2}} \right].
\end{align}

\begin{figure}[ht]
	\centering
	\includegraphics[width=0.63\textwidth]{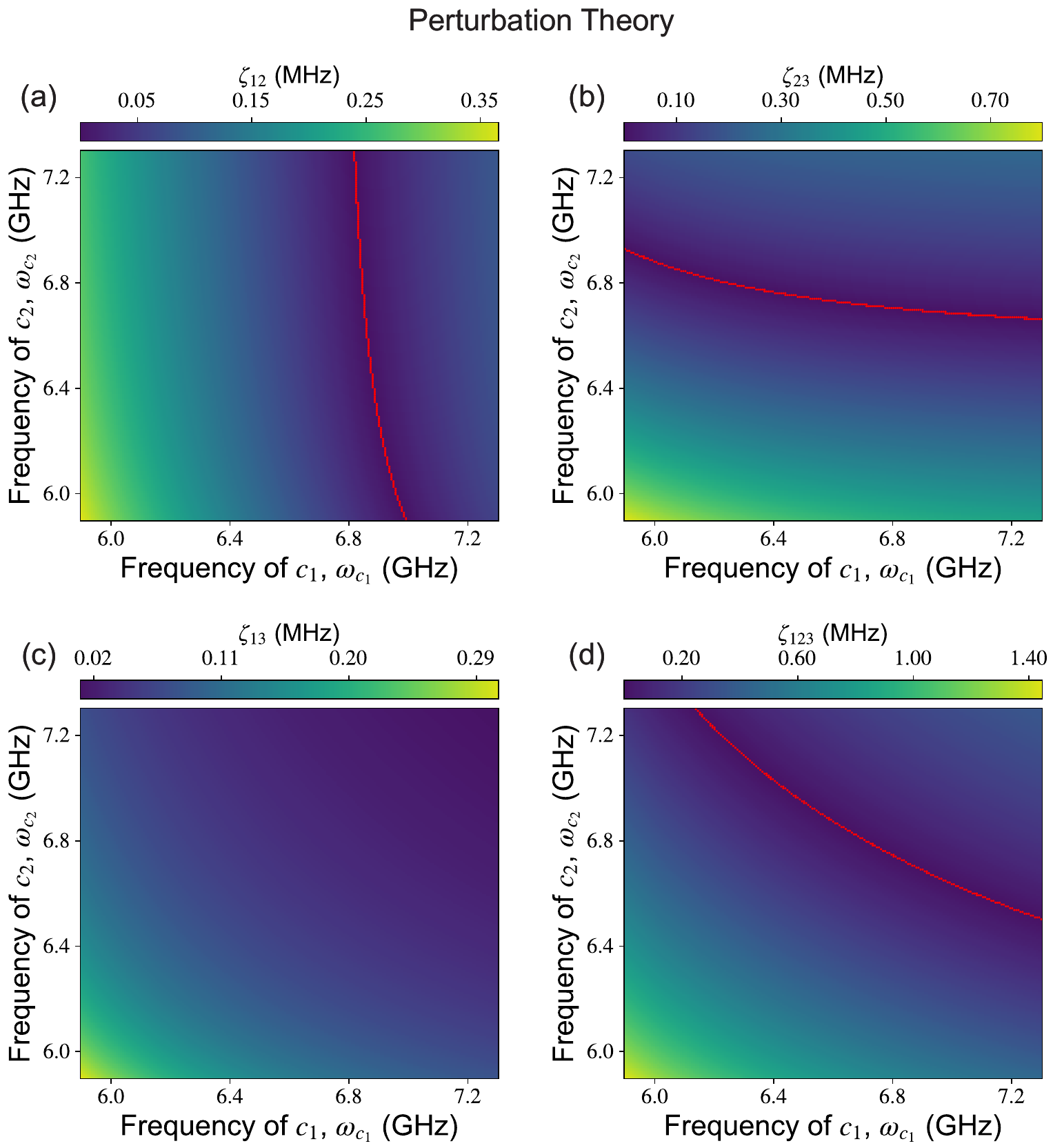}
	\caption{Calculation for the diagonal interactions from perturbation theory in the large-detuning regime. The red solid line indicates points where the couplings approach zero. (a), (b) show that $\zeta_{12}$ and $\zeta_{23}$ are mainly influenced by $\omega_{c_1}$ and $\omega_{c_2}$, respectively. (c), (d) illustrate that $\zeta_{13}$ and $\zeta_{123}$ are affected by both coupler frequencies. With the coupler frequencies decreasing below the red solid line, $\zeta_{123}$ increases significantly.
	}
	\label{fig:ZZcalculation}
\end{figure}
\begin{figure}[ht]
	\centering
	\includegraphics[width=0.63\textwidth]{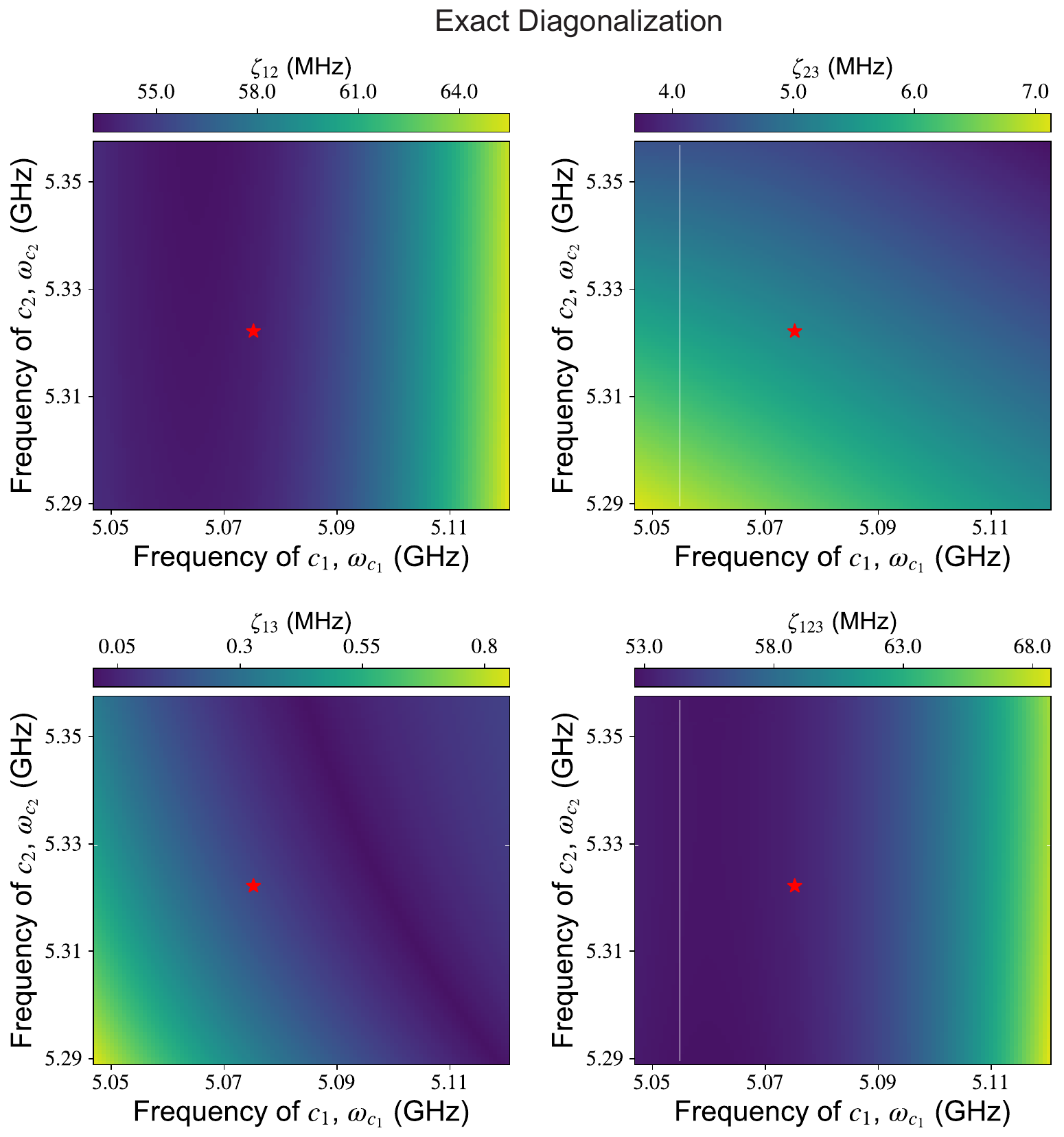}
	\caption{Calculation for the diagonal interactions via exact diagonalization of the Hamiltonian in the near-resonance regime. The optimal operating point is marked by a red star, where the coupling strengths are $\zeta_{12}=53.610~\!\mathrm{MHz}$, $\zeta_{23}=5.302~\!\mathrm{MHz}$, $\zeta_{13}=-0.157~\!\mathrm{MHz}$, $\zeta_{123}=53.411~\!\mathrm{MHz}$.}
	\label{fig:ZZdynamics}
\end{figure}

Based on the results derived from perturbation theory, we calculate the effective ZZ couplings as functions of the coupler frequencies, $\omega_{c_1}$ and $\omega_{c_2}$, in the large detuning regime. In Fig.~\ref{fig:ZZcalculation}, the red solid line indicates the points where these couplings vanish. As illustrated in Figs.~\ref{fig:ZZcalculation}(a) and (b), the couplings $\zeta_{12}$ and $\zeta_{23}$ are mainly controlled by $\omega_{c_1}$ and $\omega_{c_2}$, respectively. Meanwhile, Figs.~\ref{fig:ZZcalculation}(c) and (d) reveal that $\zeta_{13}$ and $\zeta_{123}$ depend on both $\omega_{c_1}$ and $\omega_{c_2}$. 
Notably, $\zeta_{13}$ is much smaller than the other two-body ZZ couplings, $\zeta_{12}$ and $\zeta_{23}$, indicating a weaker interaction between non-adjacent qubits. Regarding $\zeta_{123}$, we observe that as the coupler frequencies decrease below the red dashed line, the coupling strength increases significantly. This suggests a substantial enhancement of the tunable ZZZ interaction in this regime, highlighting the critical role of coupler frequency tuning for controlling multi-qubit gate operations effectively. However, the perturbation theory is no longer valid in the near-resonance regime. To accurately calculate the effective ZZ and ZZZ couplings in this regime, we employ exact diagonalization of the Hamiltonian~\cite{fors2024comprehensiveexplanationzzcoupling}, as depicted in Fig.~\ref{fig:ZZdynamics}.

\section{\label{sec:model_ccz} Three-excitation manifold}

\begin{figure}[t]
	\includegraphics[width=0.96\textwidth]{ 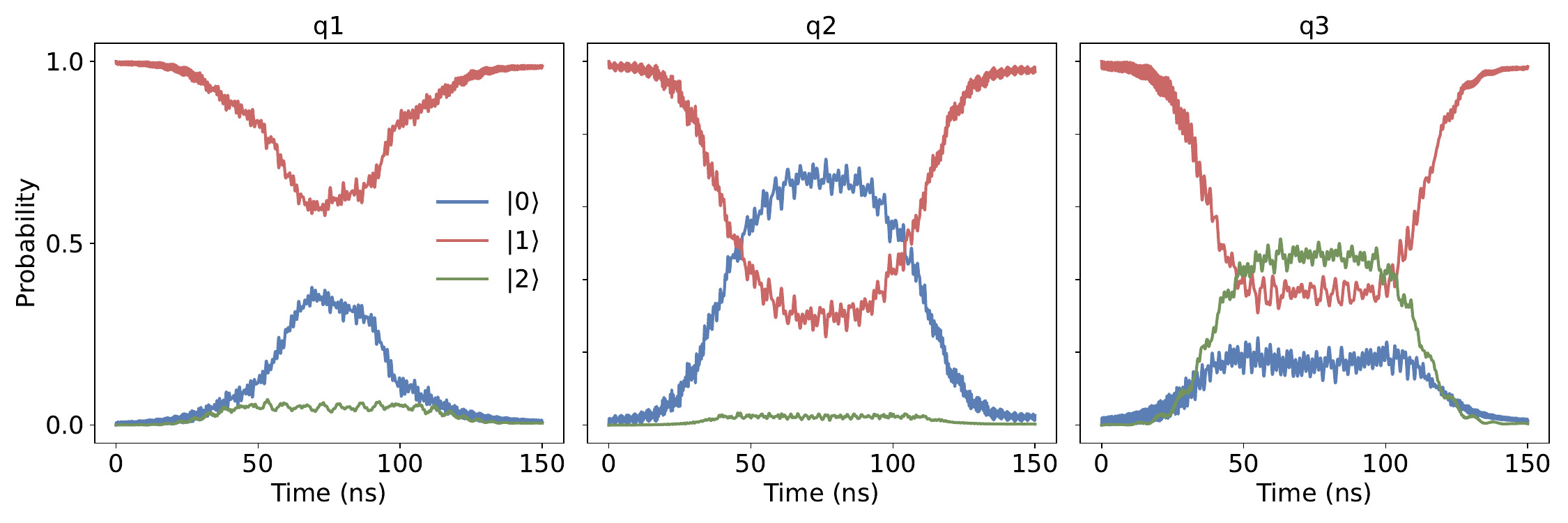}
	\caption{Evolution of state probabilities within the three-excitation manifold.
    }\label{fig:012}
\end{figure}

\begin{figure}[ht!]
	\includegraphics[width=0.96\textwidth]{ 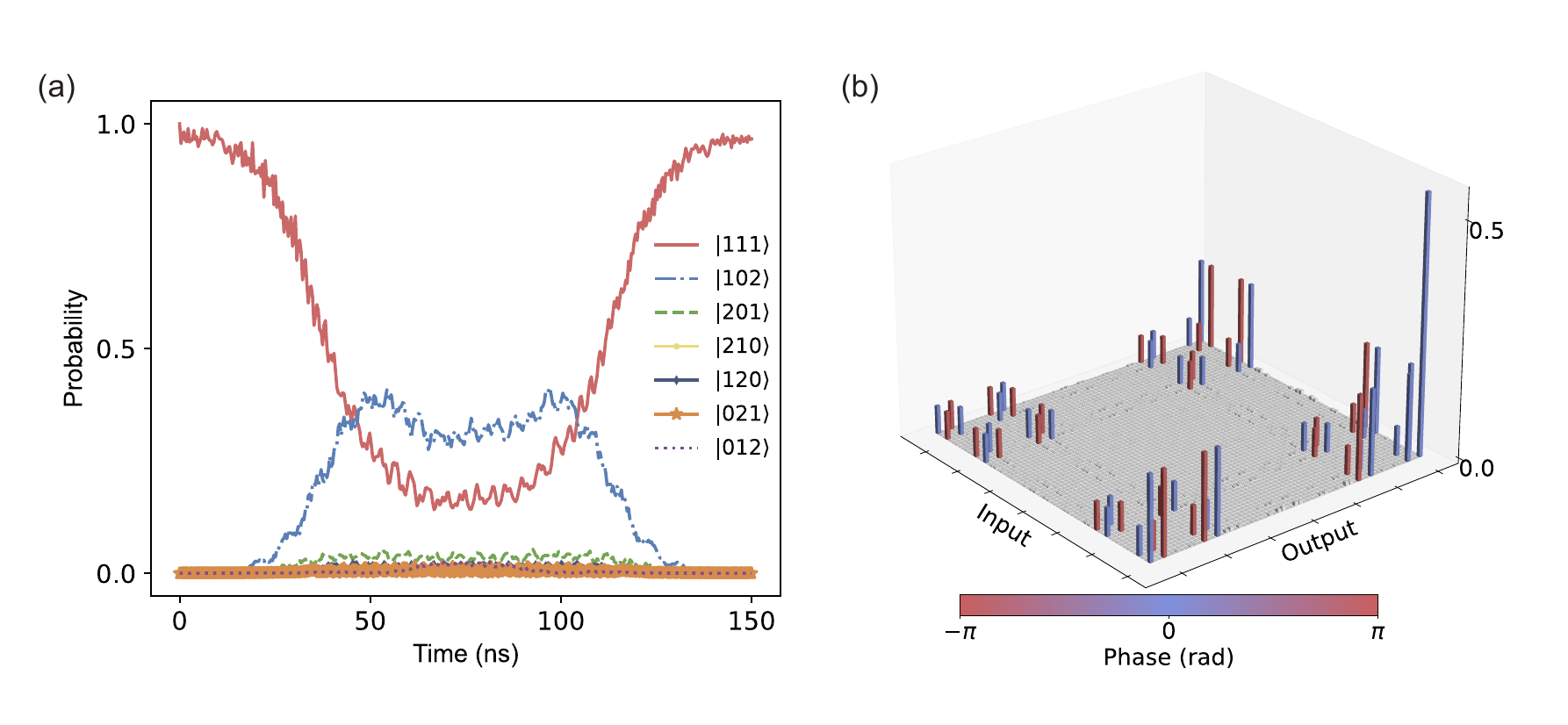}
	\caption{Numerical simulation of CCZ gate with $g_{13}=0$. (a) The time evolution probabilities of the three-excitation states. (b) The $\chi$-matrix for 64 initial states.
    }\label{fig:g13_0}
\end{figure}

\begin{figure}[t]
	\includegraphics[width=0.96\textwidth]{ 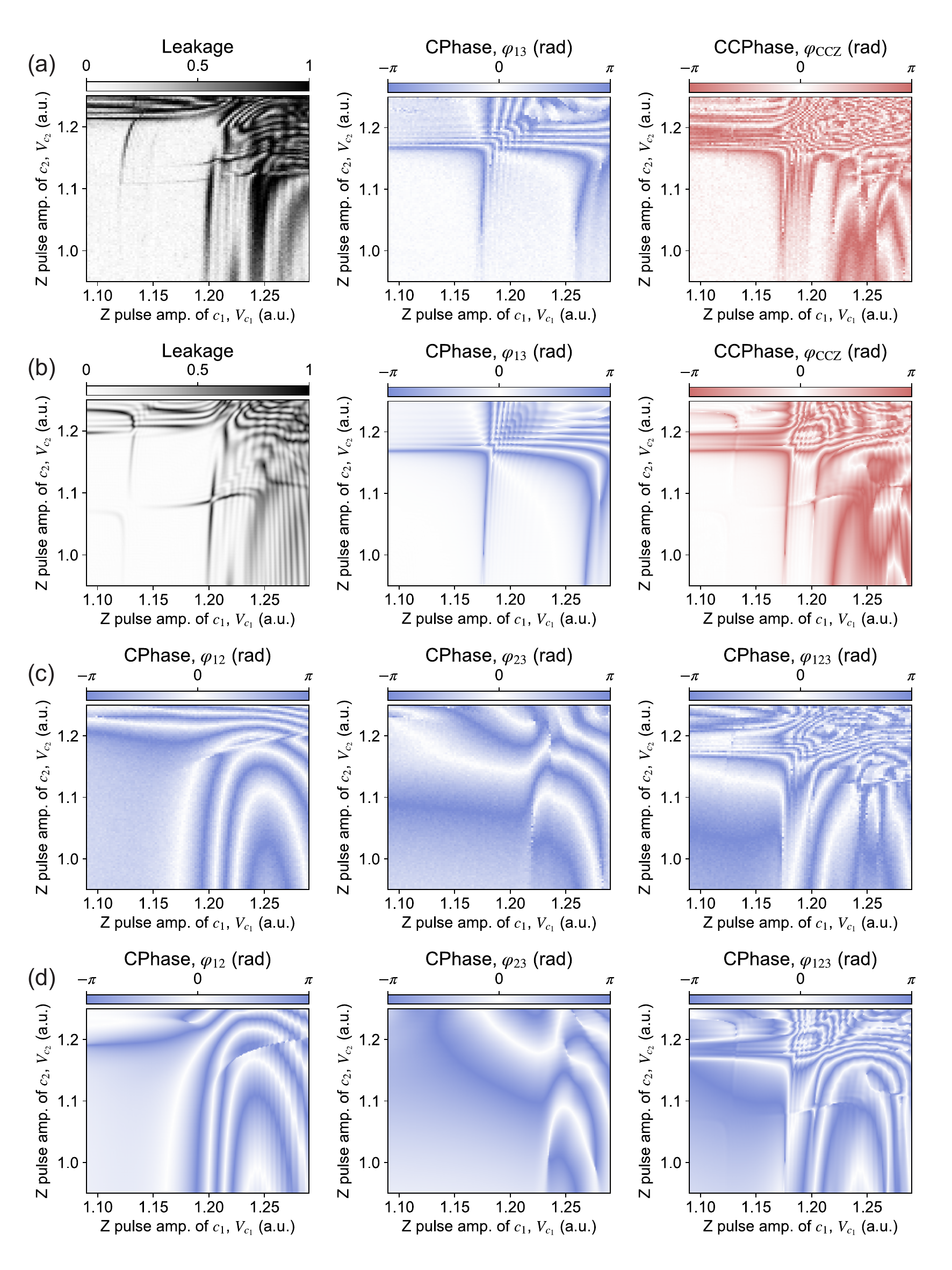}
	\caption{Tune-up measurements for the CCZ gate. 
	(a) and (c) display experimental data of leakage, two-qubit CPhases, and three-qubit CCPhase as functions of Z pulse amplitudes for $c_1$ ($x$ axis) and $c_2$ ($y$ axis). 
	(b) and (d) present numerical simulations that replicate the experimental tune-up measurements.
    }\label{fig:num_exp}
\end{figure}

The implementation of a CCZ gate requires the computational state $\ket{111}$ to undergo cyclic evolution, acquiring a conditional $\pi$-phase shift while preserving its population. In the following, we will elucidate the operational principle of the CCZ gate by analyzing the dynamics within the three-excitation subspace. In the CCZ gate experiment, as the coupler frequency is tuned downwards, the qubit frequencies decrease due to the ac-Stark effect, which may result in interactions with higher energy levels. We identify the states that interact strongly with the $|111\rangle$ within the three-excitation manifold, as shown in Fig.~\ref{fig:012}. When implementing the CCZ gate, we bring $|102\rangle$ into resonance with $|111\rangle$, thereby generating a three-qubit conditional phase. The interaction between $|102\rangle$ and $|111\rangle$ is significantly stronger than that between $|201\rangle$ and $|111\rangle$, since it is easier to achieve $\omega_2=\omega_3+\alpha_3$ compared to $\omega_2=\omega_1+\alpha_1$. Similarly, the interactions between the states $|021\rangle$, $|012\rangle$, $|120\rangle$, $|210\rangle$ with $|111\rangle$ are comparatively weak. The above explanation can be modeled as the Hamiltonian in the subspace of three excitations:
\begin{equation}
	\overset{
	\setlength{\arraycolsep}{20 pt}
	\begin{array}{ccccccc}
	\hspace{33pt} \ket{210} &  \ket{201} & \ket{120} &  \ket{111} &  \ket{102} &  \ket{021} &  \ket{012}
	\end{array}
	}
	{
	  H_3^{\mathrm{eff}} =
	\begin{pmatrix}
	\:2\omega_1 \!+\! \alpha_1 \!+\! \omega_2 & \:-g_{23} & \:-2g_{12} & \:-\sqrt{2}g_{13} & \:0 & \:0 & \:0  \\
	\:-g_{23} & \:2\omega_1 \!+\! \alpha_1 \!+\! \omega_3 & \:0 & \:-\sqrt{2}g_{12} & \:-2g_{13} & \:0 & \:0  \\
	\:-2g_{12} & \:0 & \:2\omega_2 \!+\! \alpha_2 \!+\! \omega_1 & \:-\sqrt{2}g_{23} & \:0 & \:-g_{13} & \:0 \\
	\:-\sqrt{2}g_{13} & \:-\sqrt{2}g_{12} & \:-\sqrt{2}g_{23} & \:\omega_1 \!+\! \omega_2 \!+\! \omega_3 & \:-\sqrt{2}g_{23} & \:-\sqrt{2}g_{12} & \:-\sqrt{2}g_{13} \\
	\:0 & \:-2g_{13} & \:0 & \:-\sqrt{2}g_{23} & \:2\omega_3 \!+\! \alpha_3 \!+\! \omega_1 & \:0 & \:-g_{12} \\
	\:0 & \:0 & \:-g_{13} & \:-\sqrt{2}g_{12} & \:0 & \:2\omega_2 \!+\! \alpha_2 \!+\! \omega_3 & \:-2g_{23} \\
	\:0 & \:0 & \:0 & \:-\sqrt{2}g_{13} & \:-g_{12} & \:-2g_{23} & \:2\omega_3 \!+\! \alpha_3 \!+\! \omega_2  \\
	\end{pmatrix},
	}
\end{equation}
where the time-dependent evolution is governed by $U_3^{\mathrm{eff}}=\exp({-\mathrm{i}\int_0^{\tau}H_3^{\mathrm{eff}}\mathrm{d}t})$.

Additionally, we investigate the role of $g_{13}$ in the CCZ experiment. With $g_{12}=5.0~\!\mathrm{MHz}$ and $g_{23}=6.0~\!\mathrm{MHz}$, we calculate the time-dependent probabilities of the three-excitation states for two cases: $g_{13}=0.3~\!\mathrm{MHz}$ and $g_{13}=0$, corresponding to Fig.~1(f) (in the main text) and Fig.~\ref{fig:g13_0}(a), respectively. Furthermore, we also obtain the $\chi$-matrix for 64 initial states, as illustrated in Fig.~\ref{fig:g13_0}(b). This matrix closely resembles the one presented in Fig.~\ref{fig:chi}(a), suggesting that the $g_{13}$ has minimal impact on the three-body interaction.


\section{\label{sec:data_compare} Pulse parametrization and numerical data}
The implementation of the CCZ gate requires simultaneous application of Z pulses to couplers, ensuring precise frequency control and synchronization through the conversion of these voltage pulses into magnetic flux pulses via the respective coupler Z lines. Here we adopt a flat-top Gaussian pulse shape to minimize the total duration of the first segment of the CCZ while maintaining sufficient adiabaticity to suppress energy leakage. The pulse shape is described by the following equation:
\begin{equation}
	V_c(t) = \frac{V_c}{2} \left[ \mathrm{erf}  \left( \frac{t_0+\tau-t}{\sqrt{2}\sigma} \right) - \mathrm{erf}  \left( \frac{t_0-t}{\sqrt{2}\sigma} \right) \right]\label{eq:flattop}
\end{equation}
where $V_c$ is the Z pulse amplitude, $t_{0}$ denotes the starting time of the pulse, $\tau$ represents the duration, respectively. The parameter $\sigma=50~\!\mathrm{ns}$ represents the width of the Gaussian filter.\par
Given this control waveform
, we experimentally calibrate the CCZ gate, as shown in Fig.~\ref{fig:num_exp}(a). Meanwhile, We employ time-dependent Hamiltonian simulations to replicate the experimental procedure, yielding results that closely align with the experimental observations, as illustrated in Fig.~\ref{fig:num_exp}(b). The experimental and numerical data for the conditional phases $\varphi_{123}$, $\varphi_{12}$, $\varphi_{23}$ are shown in Fig.~\ref{fig:num_exp}(c), (d), respectively. The slight discrepancies between experimental and numerical results can be attributed, partly, to the omission of dissipation effects in the numerical simulations. To account for these dissipation dynamics, we employ the Lindblad master equation
\begin{equation}
	\frac{d\rho(t)}{dt}=-\frac{i}{\hbar}[H,\rho(t)]+\sum_{n=1}^{L} \left( K_n \rho(t) K_n^{\dagger}-\frac{1}{2} \{K_n^{\dagger}K_n,\rho(t) \} \right)
	\label{eq:Lindblad}
\end{equation}
where $\rho(t)$ denotes the time-dependent density matrix of $L$ qubits, and $K_n, K_n^{\dagger}$ are  Lindblad operators. The last term in Eq.~\ref{eq:Lindblad} describes dissipation arising from the interaction of the system with the environment. In the following session, we consider decoherence effects on both the qubits and couplers. The Lindblad operators in this context are defined as
\begin{equation}
    \begin{aligned}
	K_n &=(1-b^{\dagger}_n b_n)/ \sqrt{2T_2}, \\
	K_n^{\dagger} &= b_n/ \sqrt{T_1},
	\label{eq:Lindblad_operators}
	\end{aligned}
\end{equation}
where $T_1$ and $T_2$ represent the relaxation and coherence times, respectively.



\section{\label{sec:qpt} Benchmarking of the CCZ gate }
\begin{figure}[ht!]
	\includegraphics[width=0.5\textwidth]{ 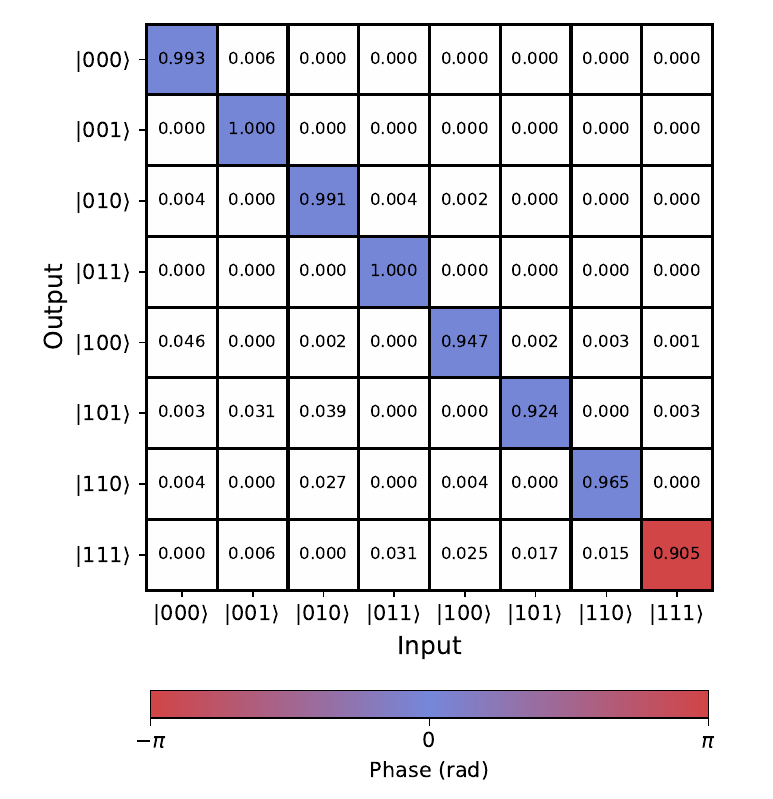}
	\caption{Truth table data for the direct CCZ gate, corresponding to Figure 3(a) in the main text.}\label{fig:88}
\end{figure}

\begin{figure}[ht!]
	\includegraphics[width=0.91\textwidth]{ 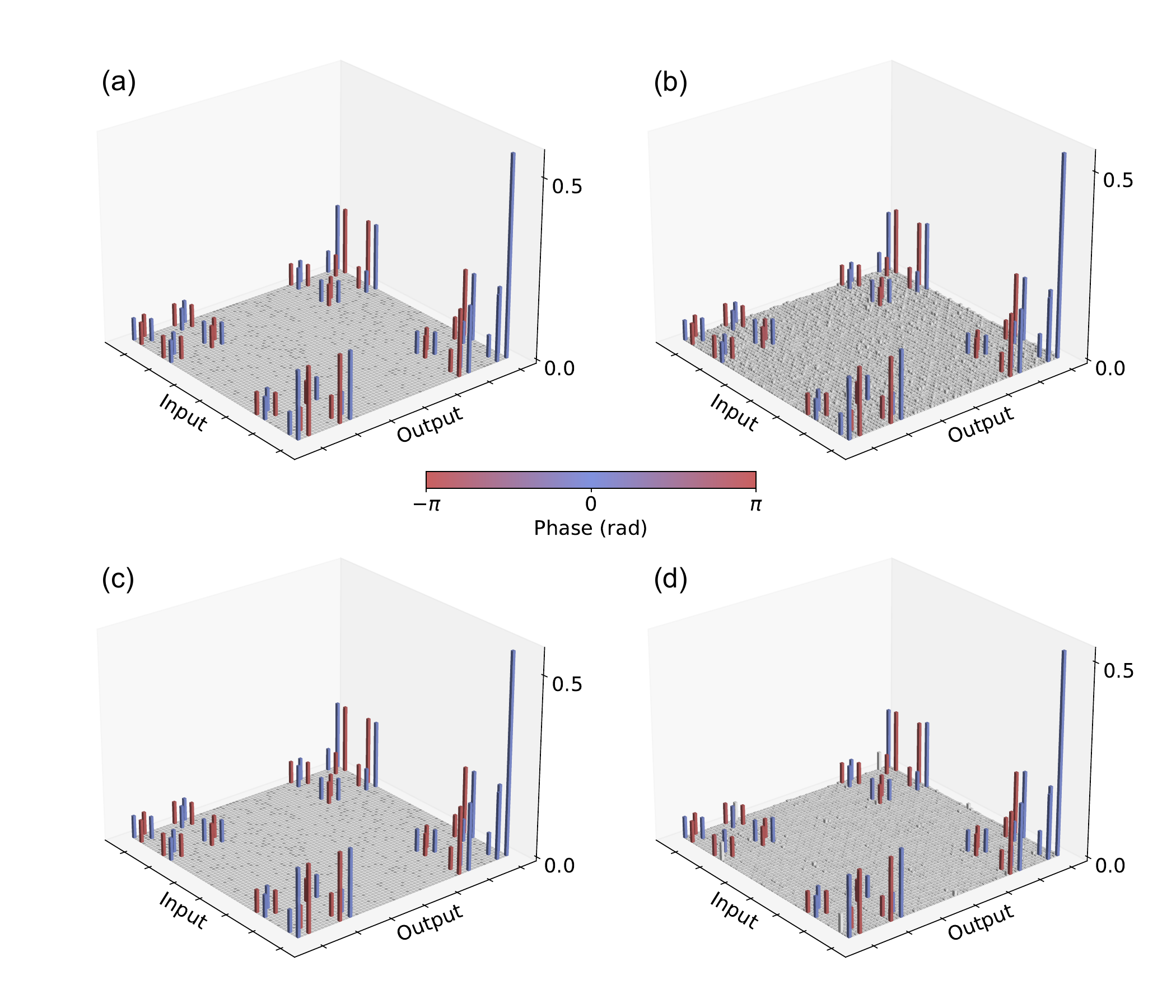}
	\caption{$\chi$-matrix for CCZ gate. (a) and (b) present $\chi_{\mathrm{ideal}}$ and $\chi_{\mathrm{exp}}$ for 64 states, respectively. 
    (c) and (d) denote $\chi_{\mathrm{ideal}}$ and  $\chi_{\mathrm{exp}}$ for 216 states, respectively.
    }\label{fig:chi}
\end{figure}

The detail of the truth table is obtained for eight computational basis states, as illustrated in Fig.~\ref{fig:88}. The visibility of the truth table is calculated as $\mathrm{Tr}[U_{\mathrm{exp}}U_{\mathrm{ideal}}]/8=96.52\%$, with an average final state fidelity of $97.94\%$. To further assess the precision of the CCZ gate, we initial three quibts into 64 and 216 states using single-qubit operations $\{I, X, X/2, Y/2\}$ and $\{I, X, X/2, -X/2, Y/2, -Y/2\}$, respectively. State tomography is applied to the output states, allowing us to obtain the density matrix and calculate both $\chi_{\mathrm{ideal}}$ and $\chi_{\mathrm{exp}}$, as shown in  Fig.~\ref{fig:chi}. The process fidelity is $93.54\%$ for the 64 states and $89.54\%$ for the 216 states. Additionally, the average final state fidelity for the 64 states is $97.07\%$, while for the 216 states, it is $96.20\%$. 

\begin{table}[htbp!]
	\begin{ruledtabular}
		\begin{tabular}{cccccc}
			Qubit & {$Q_1$} & {$Q_2$} & $Q_3$ & {$C_1$} & $C_2$\\[.02cm] \hline
	        Relaxation Time, $T_1$ ($\mathrm{\mu s}$)& 37.76& 28.15& 43.34& 27.9& 59.13\\
	        Coherence Time, $T_2$ ($\mathrm{\mu s}$)& 20.42& 20.39& 20.95& 20.45&  21.35\\
		\end{tabular}
	\end{ruledtabular}
	\caption{Decoherence parameters.}\label{tab:decoherence_paras}
\end{table}

\begin{figure}[ht!]
	\includegraphics[width=0.5\textwidth]{ 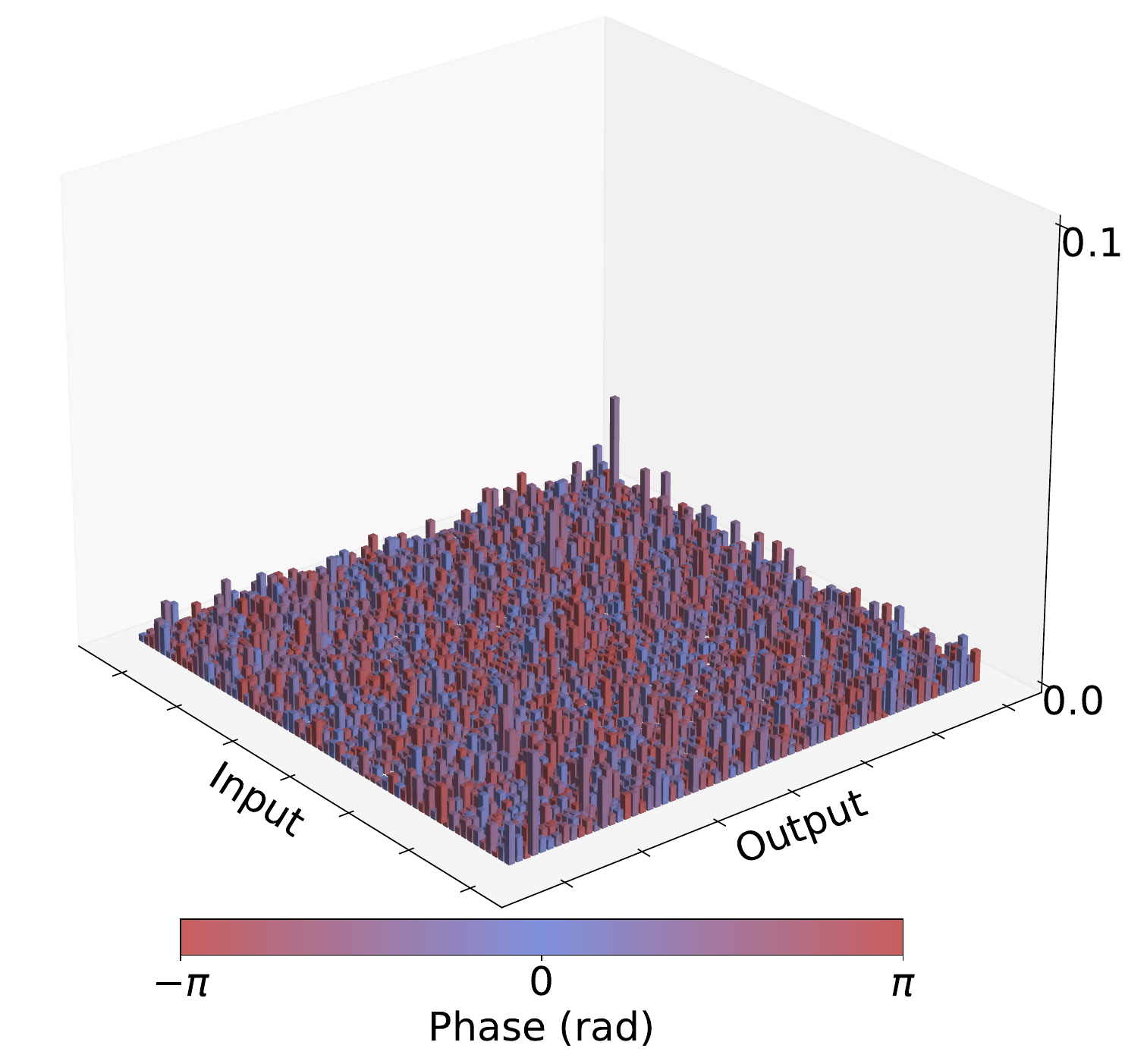}
	\caption{Difference between numerical and experimental $\chi$-matrix. The absolute value of the difference is represented by the height of the bars, with a maximum value of 0.022, while the phase of the difference is indicated by the color of the bars.
    }\label{fig:decoherence}
\end{figure}

\section{\label{sec:error} Numerical analysis of gate error}
We conduct numerical simulation to quantify the sources of error in the CCZ gate, 
and find that the decoherence of qubits is one of the most significant contributors to the error in our implementation ~\cite{PhysRevLett.129.150504,abad2023impact}. In the absence of decoherence effects, the process fidelity of the CCZ gate reaches $98.75\%$, demonstrating the high potential of our scheme for achieving a high-fidelity three-qubit gate. When considering the decoherence dissipation factor, as detailed in Table \ref{tab:decoherence_paras}, the resulting process fidelity decreases to \textbf{$93.54\%$}, closely matching the experimental results. Additionally, the $\chi$-matrices derived from both simulations and experiments exhibit strong consistency, as shown in Fig.~\ref{fig:decoherence}. The height of the bars represents the absolute difference between the two, with a maximum value of 0.022.

\section{\label{sec:decomposed} Decomposition of CCZ gate}
\begin{figure}[ht!]
	\includegraphics[width=0.8\textwidth]{ 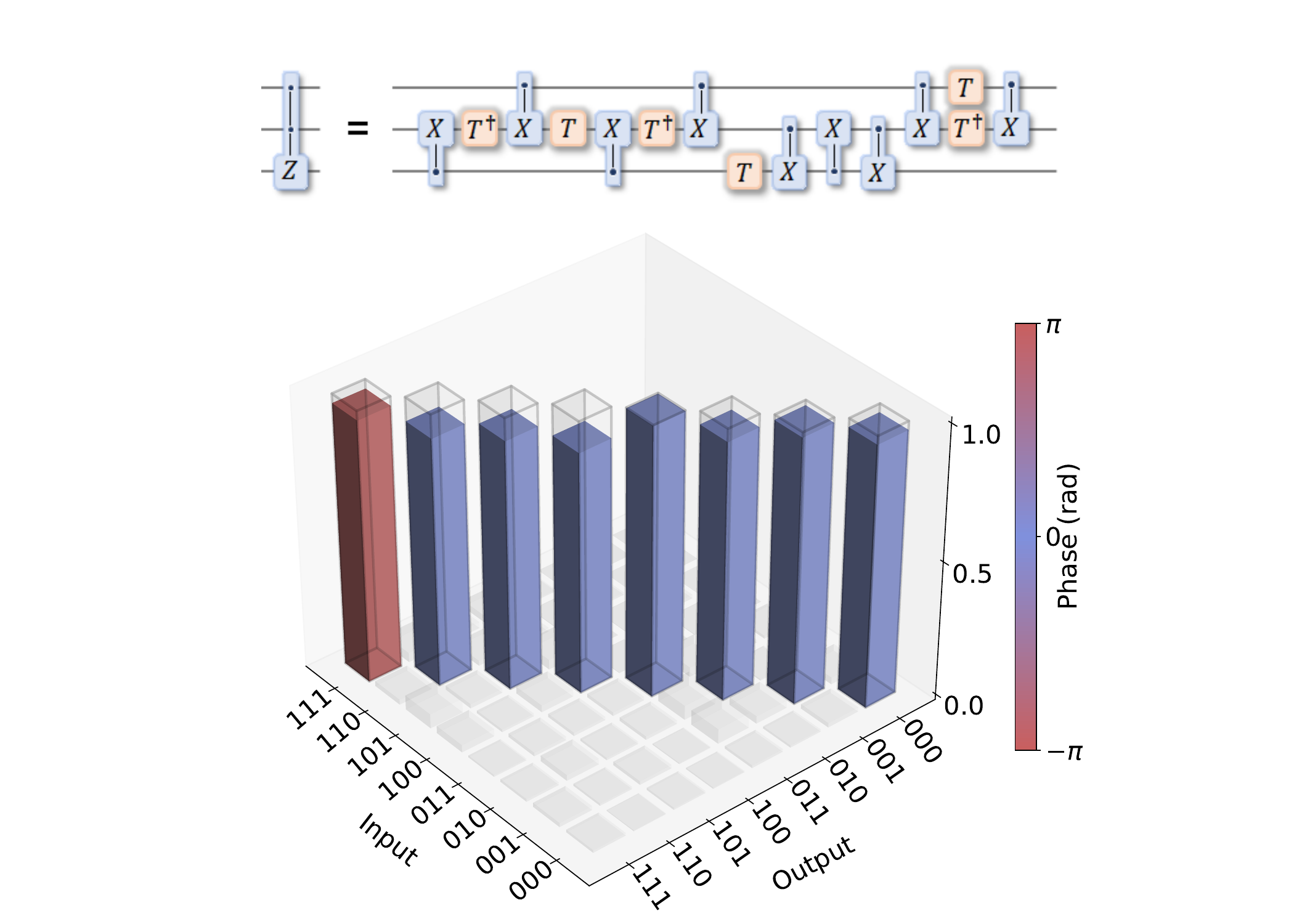}
	\caption{Experimental truth table for decomposed CCZ gate. The upper part shows the quantum circuit of the CCZ gate decomposed into a series of single-qubit gates and CX (CNOT) gates.
    }\label{fig:decomposed}
\end{figure}

\begin{figure}[ht!]
	\includegraphics[width=0.9\textwidth]{ 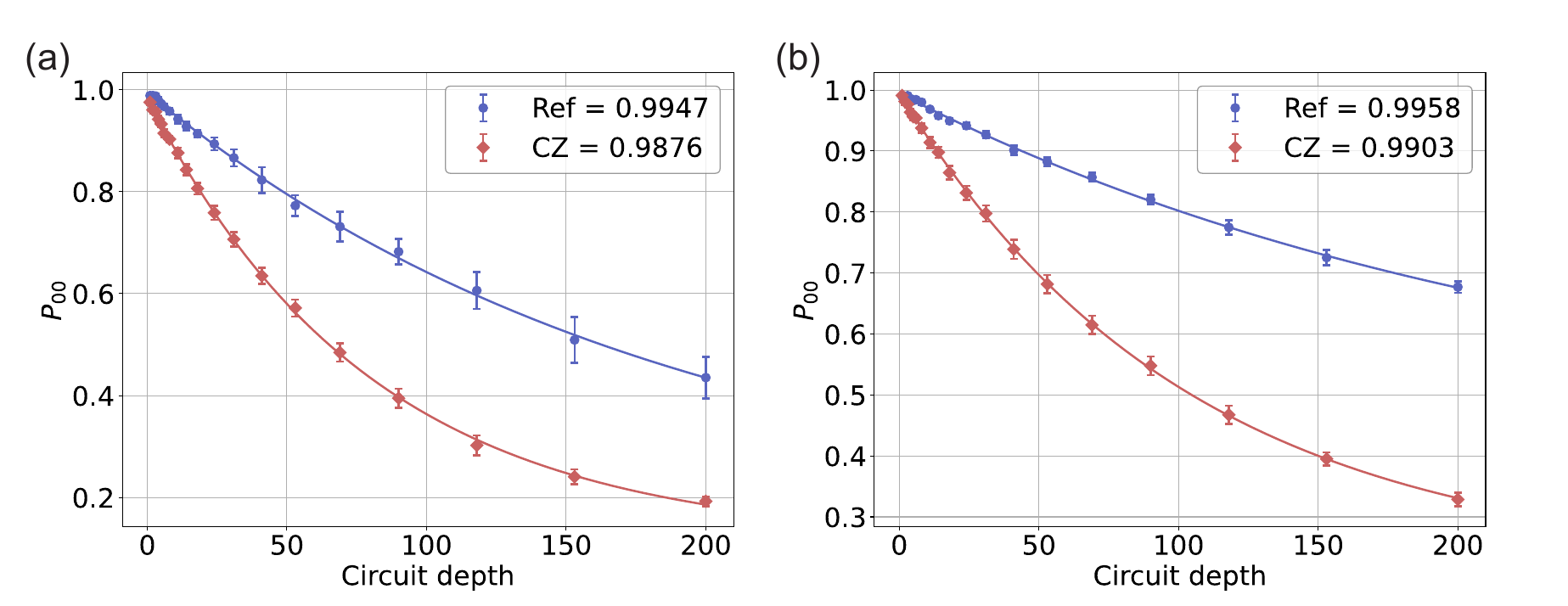}
	\caption{Experimental randomized benchmarking (RB) data for CZ gates. Subplots (a) and (b) display the RB results for the qubit pairs $q_1q_2$ and $q_2q_3$, repectively. The blue lines represent the fitting results for the reference fidelities as a function of circuit depth, while the red lines depict the fitting results for the interleaved CZ gate fidelities. The resulting RB fidelities for $q_1q_2$ and $q_2q_3$ are $99.46\%$ and $99.59\%$, repectively.
    }\label{fig:RB}
\end{figure}

The decomposed CCZ gate in our experiment is constructed by using a combination of eight CNOT gates and seven $T$ gates, with the truth table depicted in Fig.~\ref{fig:decomposed}. The visibility of the measured truth table is $94.57\%$, with an average final state fidelity of $97.15\%$.
Each CNOT gate is implemented using one CZ gate and two Hadamard gates. $T$ and $T^{\dagger}$ corresponds to the single-qubit roations $R^{Z}_{\pi/4}$ and $R^{Z}_{-\pi/4}$, respectively. To evaluate the performance of the CZ gate, we perform the randomized benchmarking (RB). As shown in Fig.~\ref{fig:RB}, for the qubit pair consisting of $q_1$ and $q_2$, we measure and obtain a reference fidelity $p_{\mathrm{ref}}$ of $99.47\%$ and a interleaved CZ fidelity $p_{\mathrm{CZ}}$ of $98.76\%$, yielding a RB fidelity of $F_{\mathrm{RB}}=1-(1-p_{\mathrm{CZ}}/p_{\mathrm{ref}})(1-1/d)=99.46\%$, where $d=4$ is the dimension of the two-qubit Hilbert space. Similarly, for the pair $q_2$ and $q_3$, the reference fidelity reaches $99.58\%$, while the CZ gate fidelity achieves $99.03\%$, resulting in a RB fidelity of $99.59\%$.

\section{\label{sec:grover} Demonstration of three-qubit Grover search algorithm}
\begin{figure}[ht!]
	\includegraphics[width=0.9\textwidth]{ 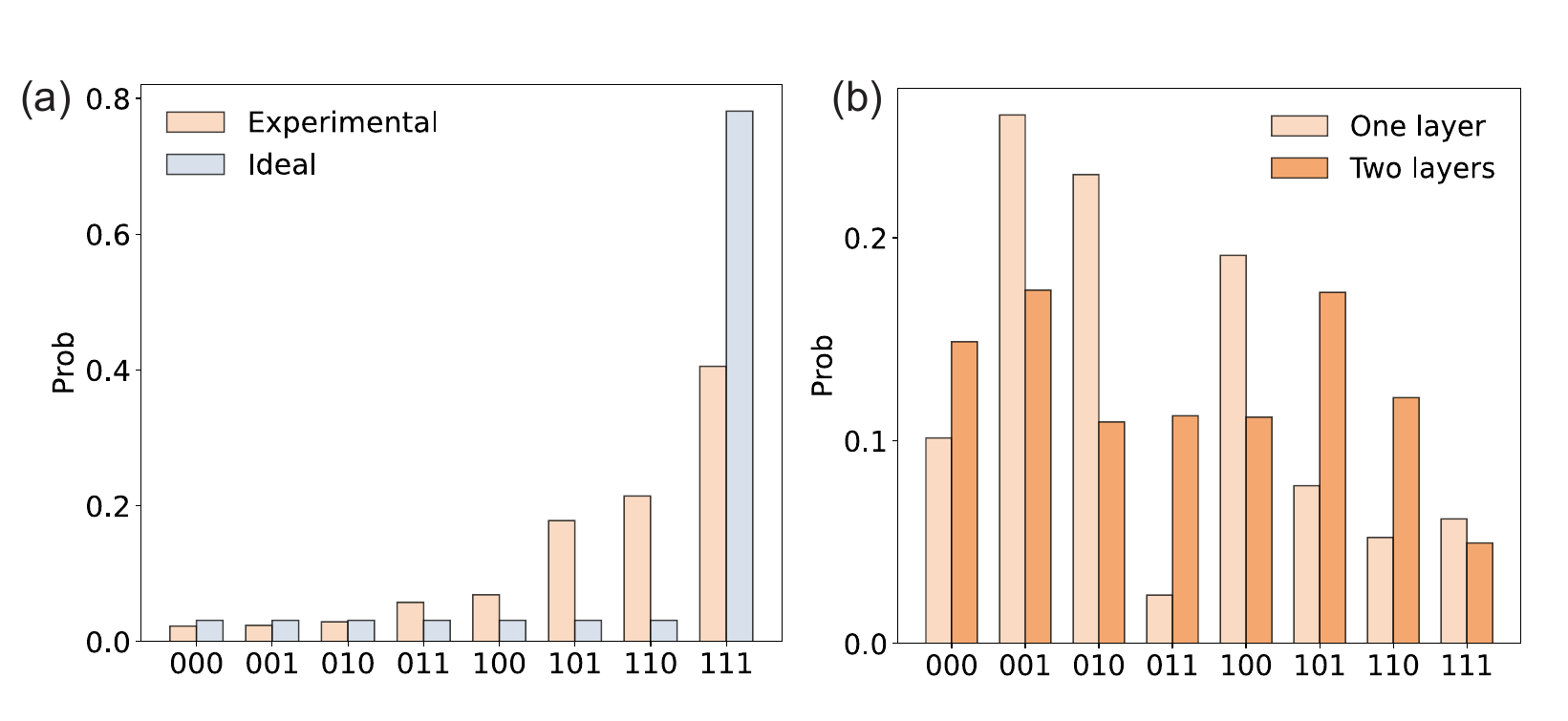}
	\caption{Joint probability data of three-qubit Grover search algorithm. (a) Experimental and ideal results after performing a single Grover operation. (b) Experimental results obtained after executing the Grover algorithm utilizing the decomposed CCZ gate.
    }\label{fig:Grover1_decomposed}
\end{figure}

In our experiment, we present the state probabilities following a single Grover operation, as shown in Fig.~\ref{fig:Grover1_decomposed}(a). Additionally, we implement the three-qubit Grover search algorithm using the decomposed CCZ gate, as shown in Fig.~\ref{fig:Grover1_decomposed}(b). The performance is notably inferior compared to that achieved with our direct CCZ gate implementation, highlighting the superior accuracy and practicality of our approach for three-qubit gates.

\clearpage
%